\documentclass[graybox]{svmult}

\usepackage{mathptmx}       
\usepackage{helvet}         
\usepackage{courier}        
\usepackage{type1cm}        
\usepackage{makeidx}         
\usepackage{graphicx}        
\usepackage{multicol}        
\usepackage[bottom]{footmisc}

\makeindex             
                       
\usepackage[numbers, sort, compress]{natbib}                       
\usepackage{amsmath}
\usepackage{amssymb}
\usepackage{bbm}

\usepackage{enumerate}

\usepackage{units}

\usepackage{color}
\usepackage{url}

\usepackage[colorlinks]{hyperref}
\hypersetup{%
        plainpages=true,
        breaklinks=true,
        hypertexnames=false,
        pageanchor=true,
        colorlinks=true,
        linkcolor={blue},
        citecolor={magenta},
        urlcolor={blue},
        anchorcolor={black}
      }

\usepackage{mleftright} 

\newcommand{\figref}[1]{\mbox{Fig.~\ref{#1}}}

\newcommand{\secref}[1]{\mbox{Sec.~\ref{#1}}}

\renewcommand{\eqref}[1]{\mbox{Eq.~(\ref{#1})}}

\newcommand{\Z}{\ensuremath{\mathbbm{Z}}}

\newcommand{\rd}{\ensuremath{\mathrm{d}}}
\newcommand{\id}{\ensuremath{\,\rd}}

\newcommand{\ket}[1]{\mleft|#1\mright\rangle}

\newcommand{\ketbra}[2]{\mleft| #1 \rangle \langle #2 \mright|}

\newcommand{\comm}[2]{\mleft[ #1, #2 \mright]}

\newcommand{\sz}{\sigma_z}
\newcommand{\sx}{\sigma_x}
\newcommand{\sm}{\sigma_-}
\renewcommand{\sp}{\sigma_+}

\newcommand{\abs}[1]{\mleft|#1\mright|}

\newcommand{\abssq}[1]{\mleft| #1 \mright|^2}

\newcommand{\nn}{\nonumber}

\newcommand{\be}{\begin{equation}}
\newcommand{\ee}{\end{equation}}
\newcommand{\bea}{\begin{eqnarray}}
\newcommand{\eea}{\end{eqnarray}}

\begin{document}

\title*{Quantum bits with Josephson junctions}

\author{Anton Frisk Kockum and Franco Nori}
\institute{Anton Frisk Kockum \at RIKEN, Saitama 351-0198, Japan \email{anton.frisk.kockum@gmail.com}
\and Franco Nori \at RIKEN, Saitama 351-0198, Japan \email{fnori@riken.jp}}

\maketitle


\section{Introduction}
\label{sec:Intro}

Already in the first edition of this book~\cite{Barone1982}, a great number of interesting and important applications for Josephson junctions were discussed. In the decades that have passed since then, several new applications have emerged. This chapter treats one such new class of applications: quantum optics and quantum information processing (QIP) based on superconducting circuits with Josephson junctions. At the time of writing, the most recent and comprehensive reviews of this field, which has grown rapidly in the past two decades, are Refs.~\cite{Gu2017, Wendin2017}. We also recommend the reviews in Refs.~\cite{You2005, Wendin2007, Clarke2008, Schoelkopf2008, You2011, Buluta2011, Zagoskin2011, Xiang2013, Devoret2013} for additional perspectives on the field. In this chapter, we aim to explain the basics of superconducting quantum circuits with Josephson junctions and demonstrate how these systems open up new prospects, both for QIP and for the study of quantum optics and atomic physics.


\subsection{What is a qubit?}
\label{sec:WhatIsAQubit}

\begin{figure}
\centering
\includegraphics[width=0.75\linewidth]{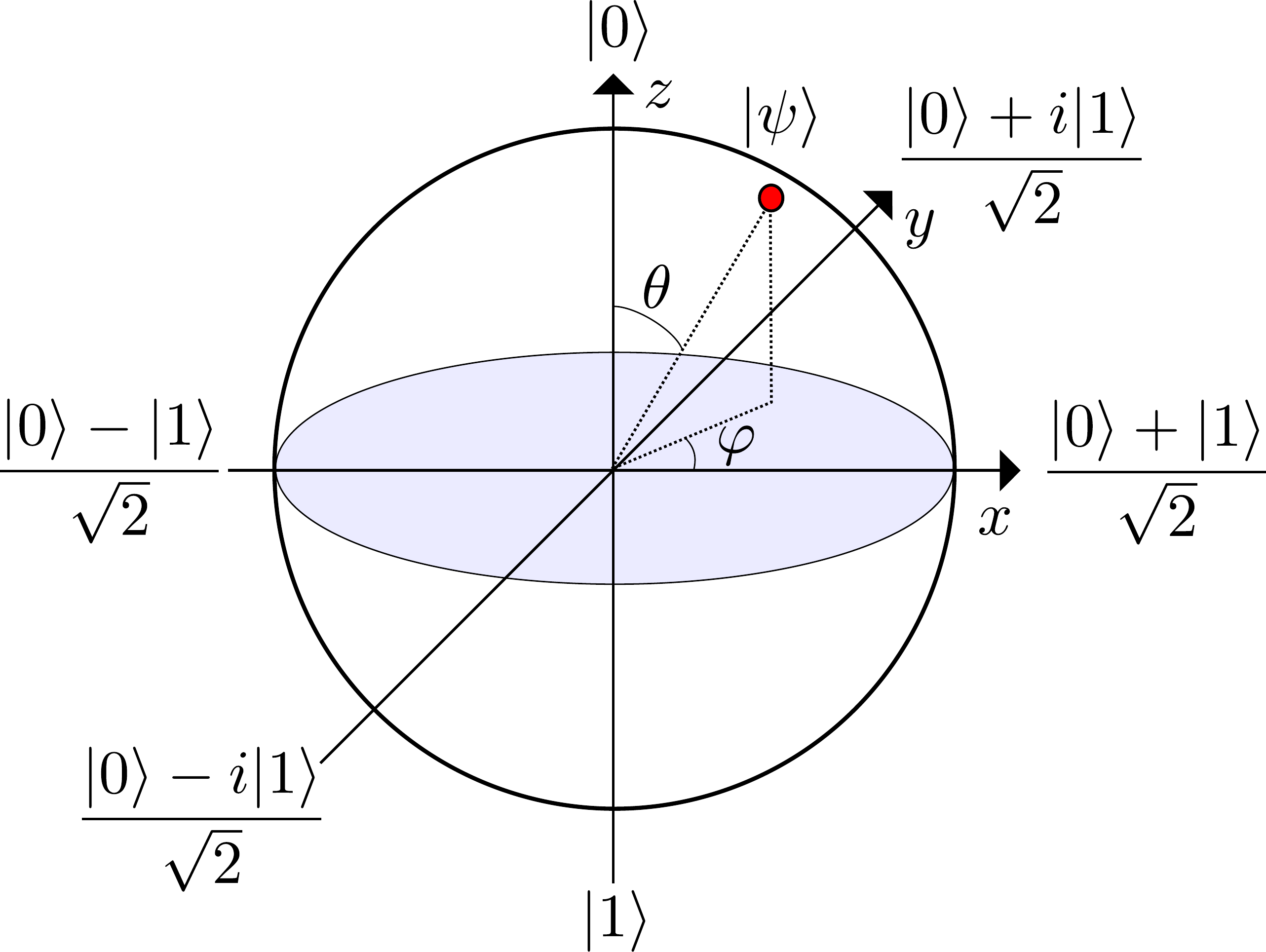}
\caption{The Bloch sphere representation of a qubit state. The north pole is the ground state $\ket{0}$ and the south pole is the excited state $\ket{1}$. To convert an arbitrary superposition of $\ket{0}$ and $\ket{1}$ to a point on the sphere, the parametrization $\ket{\psi} = \cos \frac{\theta}{2} \ket{0} + e^{i \varphi} \sin \frac{\theta}{2} \ket{1}$ is used.
\label{fig:BlochSphere}}
\end{figure}

As the name suggests, the field of QIP is concerned with information in quantum rather than classical systems. In a classical computer, the most basic unit of information is a \textit{bit}, which can take two values: 0 and 1. In a quantum computer, the laws of quantum physics allow phenomena like superposition and entanglement. When discussing information processing in a quantum world, the most basic unit is therefore a \textit{quantum bit}, usually called \textit{qubit}, a two-level quantum system with a ground state $\ket{0}$ and an excited state $\ket{1}$. Unlike a classical bit, which only has two possible states, a quantum bit has infinitely many states: all superpositions of $\ket{0}$ and $\ket{1}$,
\be
\ket{\psi} = \alpha \ket{0} + \beta \ket{1},
\ee
where $\alpha$ and $\beta$ are complex numbers satisfying $\abssq{\alpha} + \abssq{\beta} = 1$. A useful tool for visualizing a qubit state is the \textit{Bloch sphere} shown in \figref{fig:BlochSphere}. A state of the qubit is represented as a point on the surface of the sphere, which has radius 1. The two states of a classical bit correspond to the north and south poles on the sphere.

If there are $N$ qubits in a system, the total state of that system can be a superposition of $2^N$ different states: $\ket{000 \ldots 00}$, $\ket{100 \ldots 00}$, $\ket{010 \ldots 00}$, $\ldots$, $\ket{111 \ldots 10}$, $\ket{111 \ldots 11}$. This means that at least $2^N$ classical bits are required to represent this quantum system. The beginning of the field of QIP is often traced back to a talk by Feynman in 1982~\cite{Feynman1982}, where he argued for using quantum rather than classical bits to simulate quantum systems and thus achieving an exponential gain in computing resources. This would open up new avenues in, e.g., chemistry, pharmaceutics, and materials science.

Following Feynman's insight, the potential for speed-ups of computer algorithms through the use of qubits has been much studied. It has been shown that such \textit{quantum algorithms} can speed up factorization~\cite{Shor1994} (the hardness of which underpins most cryptography today), database search~\cite{Grover1997}, the solving of systems of linear equations~\cite{Harrow2009}, and several other important applications~\cite{Montanaro2016}. Note that these speed-ups are \textit{not} due to a quantum computer exploring many of the states in a superposition at the same time, but rather due to algorithms setting up interference between the complex probability amplitudes of these states in a clever way that leads to the sought answer. For a more in-depth description of the theory of quantum computation, see, e.g., the textbook in Ref.~\cite{Nielsen2000}.


\subsection{Why Josephson-junction qubits?}
\label{sec:WhyJJQubits}

To turn the enticing idea of QIP into reality, a physical implementation of qubits is needed. One option is to use single atoms or ions, well-known quantum systems. However, these tiny systems come with parameters already fixed by nature and can be hard to control. Some research groups therefore turned to circuits that can be fabricated on a chip just like the processors in today's classical computers. In addition to making fabrication relatively easy, such electrical circuits make it possible to \textit{design} the parameters of the qubits to a much greater extent, and sometimes also to \textit{tune} these parameters \textit{in situ} during an experiment. These circuits are sometimes referred to as \textit{artificial atoms}.

The superposition state of a qubit is a fragile thing, sensitive to losses. By making the circuits out of superconducting material and operating them at temperatures below the critical temperature $T_c$, resistive losses are avoided. 

\begin{figure}
\centering
\includegraphics[width=\linewidth]{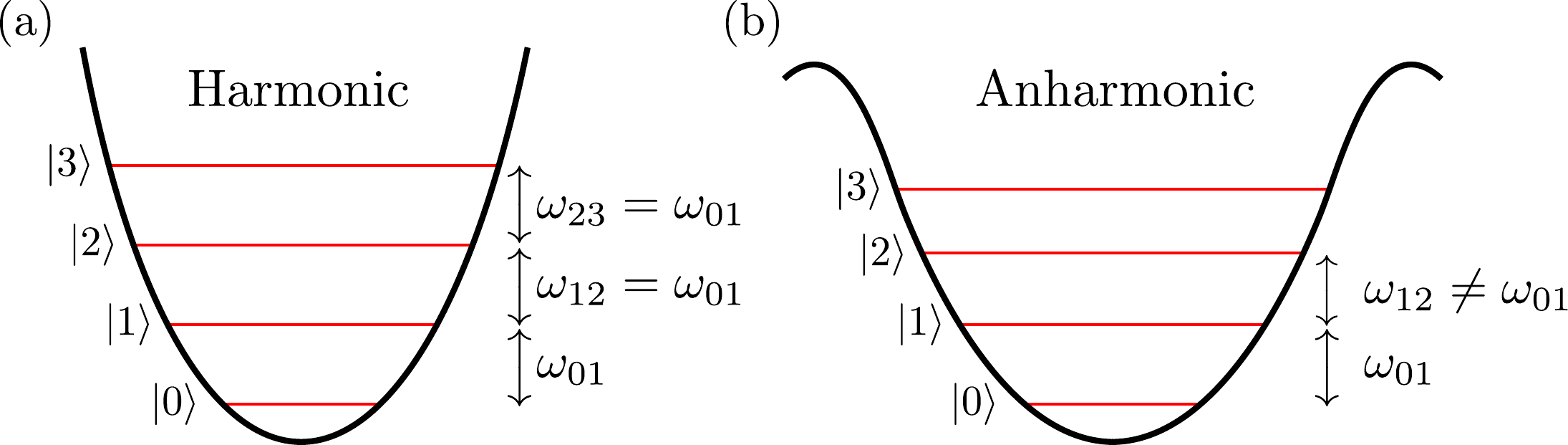}
\caption{Harmonic and anharmonic systems and their suitability as qubits.
(a) In the quadratic potential (black curve) of a harmonic system, the energy levels (red lines) are equally spaced, i.e., $\omega_{j, j+1} = \omega_{01}$, where $\omega_{jk}$ is the transition frequency between energy levels $j$ and $k$. A signal at frequency $\omega_{01}$ will thus not only transfer population from $\ket{0}$ to $\ket{1}$, but also from $\ket{1}$ to $\ket{2}$, etc.
(b) In the potential of an anharmonic system, e.g., the cosine potential characteristic of a Josephson junction, $\omega_{01} \neq \omega_{12}$. A signal at frequency $\omega_{01}$ will thus only drive transitions between $\ket{0}$ to $\ket{1}$ and not affect any other levels in the system (provided that the signal is not too strong). This limits the dynamics to the two-level system formed by $\ket{0}$ and $\ket{1}$, which can be interpreted as a qubit.
\label{fig:LinearNonLinear}}
\end{figure}

However, the superconducting circuit also needs a \textit{nonlinear element} to function as a qubit. To understand this, consider an $LC$ resonator. Such a circuit is a harmonic oscillator with resonance frequency $\omega_{\rm r} = 1/\sqrt{LC}$. When operated at low temperatures $T$ such that $\hbar \omega_{\rm r} \gg k_B T$, i.e., when thermal noise does not significantly affect quantum coherence in the system, this circuit can be treated as a quantum harmonic oscillator. As shown in \figref{fig:LinearNonLinear}(a), this quantum system has equally spaced energy levels, i.e., the energy it takes to excite the system from its ground state $\ket{0}$ to its first excited state $\ket{1}$ is the same as that required to excite the system further from $\ket{1}$ to $\ket{2}$, and so on. This means that the $LC$ resonator is not a good qubit, because when we seek to manipulate its state $\ket{\psi}$ by sending in energy at the resonance frequency, we will \textit{also} excite higher states ($\ket{2}$ and above) outside our computational subspace, which is spanned by $\ket{0}$ and $\ket{1}$.

The Josephson junction is the element that provides the nonlinearity needed to turn a superconducting circuit into a qubit. As we will see in more detail in the next sections, a Josephson junction can be incorporated into circuits in different ways to make a qubit. In a circuit where the Josephson-junction contribution dominates, the potential will be a cosine function, unlike the quadratic potential of a harmonic oscillator. As shown in \figref{fig:LinearNonLinear}(b), the spacing between energy levels in this cosine potential is anharmonic, i.e., the energy it takes to excite the system from $\ket{0}$ to $\ket{1}$ is \textit{different} from that required to excite the system from $\ket{1}$ to $\ket{2}$. This makes it possible to address the $\ket{0} \leftrightarrow \ket{1}$ transition separately to manipulate the qubit state. These manipulations can be visualized as rotations on the Bloch sphere in \figref{fig:BlochSphere}.

Josephson junctions are also an integral part of many devices needed to read out and control superconducting qubits, e.g., amplifiers, mixers, beam-splitters, switches, etc.~\cite{Gu2017}. This great reliance on Josephson junctions sets constraints on the operating temperature and frequency of the superconducting circuits discussed in this chapter. In general, the Josephson-junction qubits have transition frequencies in the range \unit[1-10]{GHz}, since this is well below the plasma frequency of the Josephson junctions involved and also matches well with frequency ranges for commercially available electronics. To ensure that $T \ll T_{\rm c}$ and $\hbar \omega_{01} \gg k_B T$, the Josephson-junction qubits are operated at temperatures on the order of \unit[10]{mK}, which is well within reach of modern dilution refrigerators.


\subsection{Outline}

In the rest of this chapter, we will further explore the world of Josephson-junction qubits. To enable a deeper understanding of how these circuits work, we first review, in \secref{sec:QuantizingCircuits}, how to quantize electrical circuits, i.e., how to derive the Hamiltonian governing their dynamics. We then apply this quantization procedure in \secref{sec:ThreeQubits} to derive the Hamiltonians for three basic types of Josephson-junction qubits: charge qubits, flux qubits, and phase qubits. In \secref{sec:FurtherQubits}, we show how these three basic types have been developed and refined further in various ways to create some of the qubits that are mainly used today. Having developed this strong foundation in the workings of Josephson-junction qubits, we then turn to their use for QIP in \secref{sec:QC}. Finally, we also discuss in \secref{sec:QO} how the artificial atoms, that Josephson-junction qubits are, have been used to explore new regimes of quantum optics and atomic physics that were hard or impossible to reach with natural atoms.


\section{Quantizing electrical circuits}
\label{sec:QuantizingCircuits}

The process for quantizing electrical circuits is briefly the following: write down the classical \textit{Lagrangian} for the circuit, identify generalized coordinates and momenta in the circuit, use these together with the Lagrangian to arrive at the \textit{Hamiltonian}, and promote the generalized coordinates and momenta to operators obeying canonical commutation relations. This process is well described in Refs.~\cite{Yurke1984, Devoret1997, Vool2017}. In this section, we cover the main points that are needed to derive Hamiltonians for the most basic Josephson-junction qubits. The material presented here and in the following two sections is mainly based on Refs.~\cite{Gu2017, Kockum2014a}.

An electrical circuit can be described as a number of nodes connected through circuit elements. As generalized coordinates for such a circuit, it is often convenient to use the \textit{node fluxes}
\be
\Phi_n (t) = \int_{-\infty}^t V_n (t') \id t',
\ee
where $V_n$ denotes the node voltage at node $n$. The corresponding generalized momenta will usually, but not every time, be the \textit{node charges}
\be
Q_n (t) = \int_{-\infty}^t I_n (t') \id t',
\ee
where $I_n$ denotes node current. However, it should be remembered that Kirchhoff's laws can reduce the number of degrees of freedom in the circuit. For example, if there is a loop $l$ in the circuit, the voltage drop around that loop should be zero, which implies
\be
\sum_{b \: \text{around} \: l} \Phi_{\rm b} = \Phi_{\rm ext},
\ee
where $\Phi_{\rm ext}$ is the external magnetic flux through $l$ and $\Phi_{\rm b}$ are the branch fluxes (not the node fluxes) around $l$. The external magnetic flux is constrained by the quantization condition $\Phi_{\rm ext} = m \Phi_0$, where $m \in \Z$ and $\Phi_0 = h / 2e$ is the flux quantum ($e$ is the elementary charge and $h$ is Planck's constant).

Once the energies of the circuit elements have been expressed in terms of the generalized coordinates $\Phi_n$ to form the Lagrangian $\mathfrak{L}$, the Hamiltonian $H$ is found by performing the \textit{Legendre transformation}~\cite{Goldstein1980}
\be
H = \sum_n \frac{\partial \mathfrak{L}}{\partial \dot{\Phi}_n} \dot{\Phi}_n - \mathfrak{L}.
\label{eq:LegendreTransf}
\ee
The ${\partial \mathfrak{L}}/{\partial \dot{\Phi}_n}$ in the first part of this expression are the generalized momenta, which often turn out to be $Q_n$.

So far, everything we have done, with the exception of the quantization condition for $\Phi_{\rm ext}$, has been classical physics. The Hamiltonian only becomes quantum when we identify the generalized coordinates and momenta as operators obeying the canonical commutation relation
\be
\comm{\Phi_n}{\frac{\partial \mathfrak{L}}{\partial \dot{\Phi}_m}} = i \hbar \delta_{nm},
\label{eq:CanonicalCommutationRelations}
\ee
where $\delta_{nm}$ is the Kronecker delta.

\begin{figure}
\centering
\includegraphics[width=0.85\linewidth]{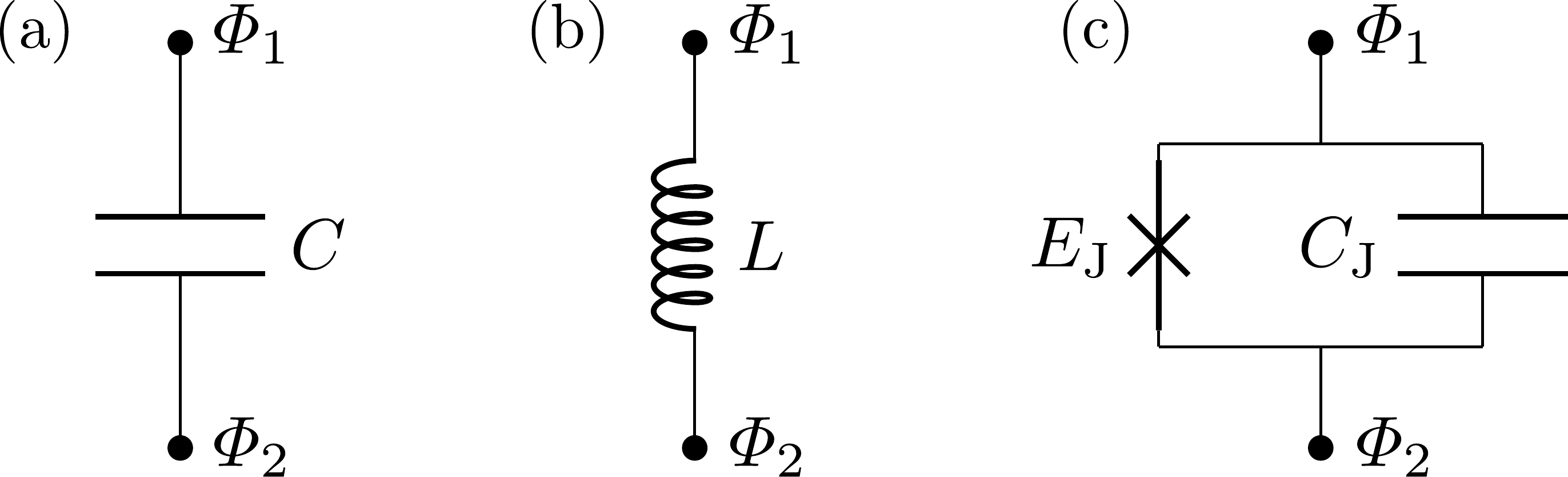}
\caption{The three basic circuit elements used to build superconducting circuits with Josephson-junction qubits.
(a) Capacitance $C$.
(b) Inductance $L$.
(c) A Josephson junction with capacitance $C_{\rm J}$ and Josephson energy $E_{\rm J}$.
\label{fig:ThreeCircuitElements}}
\end{figure}

The superconducting circuits we will discuss contain three elements: capacitors, inductors, and Josephson junctions, as shown in \figref{fig:ThreeCircuitElements}. We model the Josephson junction as a capacitor $C_{\rm J}$ in parallel with an ``X'', which contains the part characterized by the Josephson equations. The parameter needed to describe the ``X'' is the Josephson energy $E_{\rm J}$.

The Lagrangians for capacitors and inductors are easy to derive. The energy stored in a capacitor with capacitance $C$, connected to nodes with node fluxes $\Phi_1$ and $\Phi_2$ [see \figref{fig:ThreeCircuitElements}(a)], is
\be
\frac{C V^2}{2} = \frac{C \mleft( \dot{\Phi}_1 - \dot{\Phi}_2 \mright)^2}{2},
\ee
where $V$ is the voltage across the capacitor. For the case of an inductor with inductance $L$ [see \figref{fig:ThreeCircuitElements}(b)], the energy is
\be
\frac{L I^2}{2} = \mleft\{ V = L \dot{I} \mright\} = \frac{\mleft( \Phi_1 - \Phi_2 \mright)^2}{2L},
\ee
where $I$ is the current through the inductor. In the Lagrangian $\mathfrak{L}$, kinetic-energy terms give a positive contribution and potential-energy terms give negative contributions. Terms with $\dot{\Phi}$ can be identified as kinetic energy and terms with $\Phi$ correspond to potential energy. This gives
\bea
\mathfrak{L}_C &=& \frac{C \mleft(\dot{\Phi}_1 - \dot{\Phi}_2 \mright)^2}{2}, \label{eq:LagrangianC} \\
\mathfrak{L}_L &=& -\frac{\mleft( \Phi_1 - \Phi_2 \mright)^2}{2L}. \label{eq:LagrangianL}
\eea

We now turn to the Josephson junction depicted in \figref{fig:ThreeCircuitElements}(c). From the previous discussion, we already know the contribution to $\mathfrak{L}$ from the capacitive part of this circuit. To find the contribution from the "X", we recall the Josephson equations
\bea
I_{\rm J} &=& I_{\rm c} \sin \phi, \label{eq:Josephson1} \\
\dot{\phi} &=& \frac{2e}{\hbar} V (t), \label{eq:Josephson2}
\eea
where $I_{\rm J}$ is the super-current through the junction, $I_{\rm c}$ is the critical current, $V(t)$ is the voltage across the junction, and $\phi = 2e \mleft( \Phi_1 - \Phi_2 \mright) / \hbar = 2\pi \mleft( \Phi_1 - \Phi_2 \mright) / \Phi_0$ is the phase difference across the junction. Using these equations, we can calculate the energy
\be
\int_{-\infty}^t I (\tau) V (\tau) \id \tau = E_{\rm J} \mleft( 1 - \cos \phi \mright),
\ee
remembering that the Josephson energy is given by $E_{\rm J} = \hbar I_\text{c}/2e$. We can thus conclude that the Lagrangian for a Josephson junction is
\be
\mathfrak{L}_{\rm JJ} = \frac{C_{\rm J} \mleft( \dot{\Phi}_1 - \dot{\Phi}_2 \mright)^2}{2} - E_{\rm J} \mleft( 1 - \cos \phi \mright). \label{eq:LagrangianJJ}
\ee
Here we see that the cosine term enters the Lagrangian in the same way as an ordinary inductive term, i.e., it is a function of $\Phi$, not $\dot{\Phi}$. However, it is not a quadratic function of $\Phi$, which is why the Josephson junction functions as a \textit{nonlinear inductance}. As discussed in \secref{sec:WhyJJQubits}, this nonlinearity is essential for the superconducting circuits to function as qubits.

The Josephson-junction part of a superconducting qubit usually controls the transition frequency $\omega_{01}$ and other properties of the qubit. In a device with a single junction, the Josephson energy is fixed at the fabrication stage. However, by using two Josephson junctions in a SQUID configuration, a \textit{tunable} Josephson energy can be achieved, which means that various qubit parameters can be tuned during an experiment. The SQUID works as a single junction with an effective Josephson energy that is a function of the external magnetic flux through the SQUID loop~\cite{Tinkham1996}:
\be
E_{\rm J, eff} = \mleft( E_{\rm J, 1} + E_{\rm J, 2} \mright) \cos \mleft( \frac{\pi \Phi_{\rm ext}}{\Phi_0} \mright) \sqrt{1 + d^2 \tan^2 \mleft( \frac{\pi \Phi_{\rm ext}}{\Phi_0} \mright)},
\label{eq:TunableEJ}
\ee
where $E_{{\rm J}, n}$ is the Josephson energy of junction $n$ and
\be
d = \frac{E_{\rm J, 2} -  E_{\rm J, 1}}{E_{\rm J, 2} + E_{\rm J, 1}}
\ee
is a measure of the junction asymmetry.


\section{The three basic Josephson-junction qubits}
\label{sec:ThreeQubits}

\begin{figure}
\centering
\includegraphics[width=\linewidth]{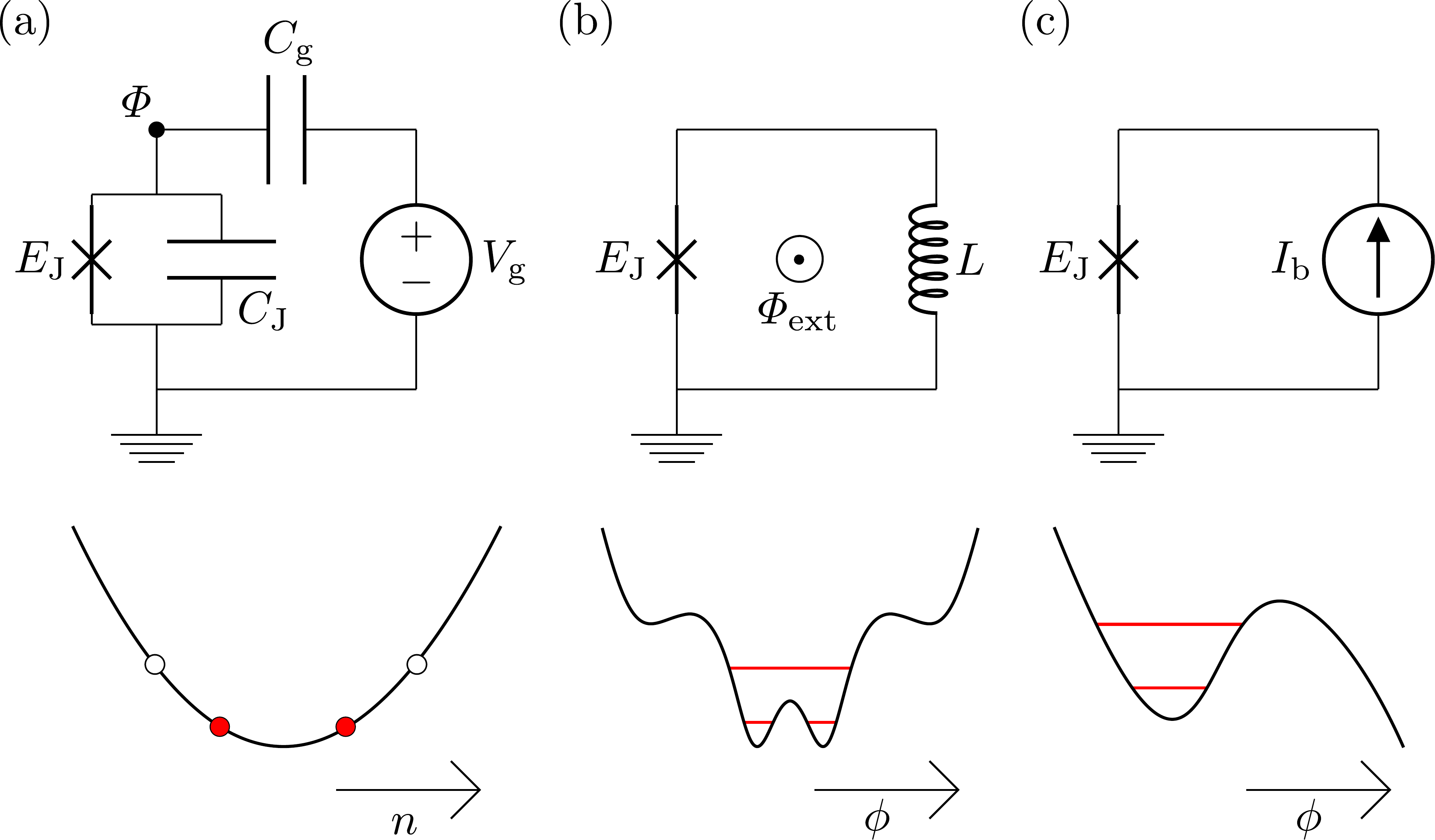}
\caption{The three basic Josephson-junction qubit circuits and their potential-energy landscapes, with the two lowest energy levels marked in red. The details of each qubit are given in the following subsections.
(a) Charge qubit. 
(b) Flux qubit.
(c) Phase qubit. For simplicity, the capacitance $C_{\rm J}$ is only shown in panel (a), although it is also present in the circuits in panels (b) and (c).
\label{fig:ThreeBasicQubits}}
\end{figure}

There are three basic designs for Josephson-junction qubits, depicted in \figref{fig:ThreeBasicQubits}. The three are known as a \textit{charge qubit} [\figref{fig:ThreeBasicQubits}(a)], a \textit{flux qubit} [\figref{fig:ThreeBasicQubits}(b)], and a \textit{phase qubit} [\figref{fig:ThreeBasicQubits}(c)], respectively. Roughly speaking, the charge qubit is a box for charge, controlled by an external voltage $V_{\rm g}$; the flux qubit is a loop controlled by an external magnetic flux $\Phi_{\rm ext}$; and the phase qubit is a Josephson junction biased by a current $I_{\rm b}$.

As we saw in \secref{sec:QuantizingCircuits}, these qubit circuits can be described by generalized coordinates and conjugate generalized momenta. If we take the phase difference $\phi$ across the Josephson junction as the coordinate, the conjugate variable will be $n$, the number of Cooper pairs on one of the superconducting islands of the junction. The commutation relation of these variables is
\be
\comm{e^{i \phi}}{n} = e^{i \phi},
\label{eq:PhiNCommutation}
\ee
which sometimes is expressed as $\comm{\phi}{n} = i$ if one does not take into account the fact that $\phi$ is periodic~\cite{Gerry2005}. From this follows that these conjugate variables obey the Heisenberg uncertainty relation $\Delta \phi \Delta n \geq 1$.

The most relevant parameter for understanding the workings of a Josephson-junction qubit is usually the ratio between the Josephson energy $E_J$ and the \textit{single-electron charging energy} $E_C = e^2 / (2C)$, where $C$ is some relevant capacitance in the circuit. When $E_J / E_C \ll 1$, the charge number $n$ is well defined and $\phi$ has large quantum fluctuations. This is the case for the charge qubit. When $E_J / E_C \gg 1$, the opposite holds. This is the case for both the flux qubit and the phase qubit.


\subsection{Charge qubit}
\label{sec:ChargeQubit}

We now discuss the three basic Josephson-junction qubits in more detail, starting with the charge qubit. The charge qubit is also known as the Cooper-pair box (CPB). It was one of the first superconducting qubits developed~\cite{Bouchiat1998, Nakamura1999, Pashkin2009}.

The upper part of \figref{fig:ThreeBasicQubits}(a) shows the circuit diagram of a CPB. The heart of the CPB is a small superconducting island (black dot with node flux $\Phi$), which is connected to a superconducting reservoir through a Josephson junction. Cooper pairs can tunnel on and off the island through this junction. The island is also connected to a voltage source $V_{\rm g}$ through a gate capacitance $C_{\rm g}$. This part of the circuit determines a background charge $n_{\rm g} = C_{\rm g} V_{\rm g} / (2e)$ (we measure the background charge in units of Cooper pairs) induced on the superconducting island by the electromagnetic environment.

We can write down the Lagrangian of the CPB circuit in \figref{fig:ThreeBasicQubits}(a) by applying Eqs.~(\ref{eq:LagrangianC}) and (\ref{eq:LagrangianJJ}):
\be
\mathfrak{L}_{\rm CPB} = \frac{ C_{\rm g} \mleft( \dot{\Phi} - V_{\rm g} \mright)^2}{2} + \frac{C_{\rm J} \dot{\Phi}^2}{2} - E_{\rm J} \mleft[ 1 - \cos \mleft( \frac{2\pi \Phi}{\Phi_0} \mright) \mright].
\ee
We then apply the Legendre transformation from \eqref{eq:LegendreTransf}, identify the conjugate momentum $Q = \mleft( C_{\rm J} + C_{\rm g} \mright) \dot{\Phi} - C_{\rm g} V_{\rm g}$ (which is the charge on the superconducting island), and remove constant terms since they do not give any contribution to the dynamics (put another way: we can set the zero energy arbitrarily; only energy differences matter). The result is the Hamiltonian
\be
H_{\rm CPB} = 4 E_{\rm C} \mleft( n - n_{\rm g} \mright)^2 - E_{\rm J} \cos \phi,
\ee
where we have identified $n = - Q / 2e$ as the number of Cooper pairs on the island and $\phi = 2e \Phi / \hbar$. Here, the capacitance defining $E_C$ is the total capacitance $C_{\rm J} + C_{\rm g}$.

Continuing to make the circuit description quantum, we promote $\Phi$ and $Q$ to operators using the commutation relation in \eqref{eq:PhiNCommutation}. From this commutation relation, it follows that $e^{\pm i\phi} \ket{n} = \ket{n \mp 1}$, where $\ket{n}$ is a system state written in the charge basis counting the number of Cooper pairs, i.e., the eigenbasis of the operator $n$. From this result, together with the resolution of unity~\cite{Sakurai1994} and the identity $\cos \phi = \mleft( e^{i \phi} + e^{- i \phi} \mright) / 2$, we obtain the CPB Hamiltonian in the charge basis:
\be
H_{\rm CPB} = \sum_n \mleft[ 4 E_{\rm C} \mleft( n - n_{\rm g} \mright)^2 \ketbra{n}{n} - \frac{1}{2} E_{\rm J} \mleft( \ketbra{n+1}{n} + \ketbra{n-1}{n} \mright) \mright].
\label{eq:HCPB}
\ee
Note that this is a tight-binding Hamiltonian with $E_{\rm C}$ and $n_{\rm g}$ determining the on-site energy and $E_J$ setting the tunneling matrix element between neighboring charge states. Since $n_{\rm g}$ can be controlled by an external voltage, it is thus possible to tune the energy levels of the system during an experiment. Further tunability, of $E_{\rm J}$, is possible if the Josephson junction is replaced by a SQUID, as explained around \eqref{eq:TunableEJ}.

The half-integer values of the background charge, $n_{\rm g} = m + \frac{1}{2}, m \in \Z$, are special due to several reasons:
\begin{itemize}
\item For these values of $n_{\rm g}$, the eigenstates of the system have \textit{well-defined parities}.
\item For the two charge states $\ket{m}$ and $\ket{m+1}$, the effective charging energies $4 E_{\rm C} \mleft( n - n_{\rm g} \mright)^2$ are \textit{degenerate} at these points.
\item At these points, the two lowest energy levels of the system are well separated from the other energy levels in the system, which makes for a good qubit. Due to the degeneracy between the charging energies of these two levels, the transition frequency for the qubit is set by $E_{\rm J}$.
\item At these points, the qubit is \textit{less sensitive} to charge noise, i.e., fluctuations in $n_{\rm g}$, since $\partial H_{\rm CPB} / \partial n_{\rm g} = 0$ here (remember that the term with $n_{\rm g}^2$ in $H_{\rm CPB}$ is a constant that can be neglected). For this reason, $n_{\rm g} = m + \frac{1}{2}$ are sometimes called \textit{sweet spots} for charge qubits.
\end{itemize}
%


\subsection{Flux qubit}

The flux qubit~\cite{Orlando1999, Mooij1999, VanderWal2000, Friedman2000, Chiorescu2003}, shown in \figref{fig:ThreeBasicQubits}(b), is another simple Josephson-junction qubit design that has been around for as long as the charge qubit. It is also known as a persistent-current qubit.

The flux qubit in its simplest form consists of a superconducting loop interrupted by one Josephson junction. However, for this circuit to function as a qubit, there must be at least two states in the local minimum of the potential energy [see the lower part of \figref{fig:ThreeBasicQubits}(b)]. Fulfilling this condition turns out to require a large self-inductance, which means that the loop needs to be large. This is not desirable when operating the circuit as a qubit, since a large loop will be more sensitive to fluctuations in external magnetic flux.

To solve the problem of inductance and loop size, the common approach is to use three Josephson junctions instead of one~\cite{Mooij1999, VanderWal2000}. Out of these three junctions, two are identical with Josephson energies $E_{\rm J}$, while the third is smaller with Josephson energy $\alpha E_{\rm J}$. The value of $\alpha$ determines the potential-energy landscape of the circuit. Usually, $\alpha$ in the range $0.6 - 0.7$ is used because it makes the circuit less sensitive to charge noise. The potential energy then looks roughly like in \figref{fig:ThreeBasicQubits}(b). However, if instead $\alpha < 0.5$, the potential energy only has a single well.

The Hamiltonian for a flux qubit with three Josephson junctions can be written as~\cite{Orlando1999}
\bea
H_{\rm flux} &=& \frac{P_{\rm p}^2}{2 M_{\rm p}} + \frac{P_{\rm m}^2}{2 M_{\rm m}} + 2 E_{\rm J} \mleft( 1 - \cos \phi_{\rm p} \cos \phi_{\rm m} \mright) \nn\\
&&+ \alpha E_{\rm J} \mleft[ 1 - \cos \mleft( 2 \pi \frac{\Phi_{\rm ext}}{\Phi_0} + 2 \phi_{\rm m}  \mright)  \mright],
\label{eq:Hflux}
\eea
with $M_{\rm p} = 2 C_{\rm J} \mleft( \Phi_0 / 2\pi \mright)^2$, $M_{\rm m} = M_{\rm p} \mleft( 1 + 2 \alpha \mright)$, $P_{\rm p} = - i \hbar \frac{\partial}{\partial \phi_{\rm p}}$, and $P_{\rm m} = - i \hbar \frac{\partial}{\partial \phi_{\rm m}}$. The phase differences across the two larger junctions, $\phi_1$ and $\phi_2$, have been combined to form the new variables $\phi_{\rm p} = \phi_1 + \phi_2$ and $\phi_{\rm m} = \phi_1 - \phi_2$.

The Hamiltonian in \eqref{eq:Hflux} can be interpreted as describing a particle with an anisotropic mass (the first two terms on the right-hand side) moving in a periodic two-dimensional potential (the last two terms on the right-hand side). Similar to how the parameters of the charge-qubit Hamiltonian in \eqref{eq:HCPB} can be tuned by changing the external voltage $V_{\rm g}$, the potential-energy term in \eqref{eq:Hflux} can be tuned by adjusting the external flux $\Phi_{\rm ext}$ (and $E_J$ can again be tuned by replacing one of the junctions with a SQUID). And just like the point $n_{\rm g} = 0.5$ is special for the charge qubit, the point $\Phi_{\rm ext} / \Phi_0 = 0.5$ is of particular interest when considering the flux qubit:
\begin{itemize}
\item At $\Phi_{\rm ext} / \Phi_0 = 0.5$, the potential-energy term is symmetric. The eigenstates of the system have well-defined parities at this point. Away from this point, the potential-energy term is asymmetric and the eigenstates no longer have well-defined parities.
\item For values of $\Phi_{\rm ext}$ such that $\Phi_{\rm ext} / \Phi_0 \approx 0.5$, the two lowest energy levels of the system are well separated from the other energy levels in the system, making the circuit a good qubit. At this point, the Hamiltonian for the two levels making up the qubit can be written as
\be
H = \frac{\varepsilon \sz + \delta \sx}{2},
\ee
where $\varepsilon = I_{\rm p} \mleft( 2 \Phi_{\rm ext} - \Phi_0 \mright)$, and the Pauli operators are defined as $\sz = \ketbra{\circlearrowleft}{\circlearrowleft} - \ketbra{\circlearrowright}{\circlearrowright}$ and $\sx = \ketbra{\circlearrowleft}{\circlearrowright} + \ketbra{\circlearrowright}{\circlearrowleft}$. Here, the basis states are $\ket{\circlearrowleft}$ and $\ket{\circlearrowright}$, i.e., states with supercurrents of magnitude $I_{\rm p}$ circulating anti-clockwise and clockwise, respectively, in the loop. Each of these circulating-current states corresponds to one potential well; the potential wells are connected by the tunneling matrix element $\delta$.
\item To first order in perturbation theory, the parameters of the flux qubit are insensitive to flux noise, i.e., fluctuations in $\Phi_{\rm ext}$. For this reason, $\Phi_{\rm ext} / \Phi_0 = 0.5$ is sometimes referred to as a \textit{sweet spot}, or optimal working point, for a flux qubit.
\end{itemize}
%


\subsection{Phase qubit}
\label{sec:PhaseQubit}

The phase-qubit circuit, depicted in \figref{fig:ThreeBasicQubits}(c), is arguably the oldest member of the Josephson-junction-qubit family. It was studied already in the 1980s as part of experimental efforts to probe quantum effects due to macroscopic degrees of freedom~\cite{Martinis1985, Clarke1988}. Further evidence that these systems are truly quantum-mechanical was later provided by demonstrating a violation of Bell's inequality~\cite{Wei2006, Ansmann2009}. These days, when Josephson-junction qubits are used for quantum computing or quantum-optics experiments, refinements of the charge and flux qubits are much more commonly seen than phase qubits, since it has turned out to be more challenging to preserve quantum coherence in the latter~\cite{Wendin2017}.

The phase qubit consists of a large Josephson junction ($E_{\rm J} / E_{\rm C} \approx 10^6$) controlled through an applied bias current $I_{\rm b}$~\cite{Martinis2009}. The bias current sets the tilt of the ``tilted-washboard'' potential for the circuit [see the lower part of \figref{fig:ThreeBasicQubits}(c)] and is usually tuned close to the critical current $I_{\rm c}$. The Hamiltonian of the circuit is
\be
H_{\rm phase} = \frac{2 \pi}{\Phi_0} \frac{p^2}{2 C_{\rm J}} - \frac{\Phi_0}{2 \pi} I_{\rm b} \phi - E_{\rm J} \cos \phi,
\ee
where the ``momentum'' $p$ is given by the charge $Q = 2e p / \hbar$ on the capacitance of the Josephson junction. Quantization proceeds as before by treating $\phi$ as the coordinate conjugate to this momentum. The resulting eigenenergies of the system have small anharmonicity, but a qubit can be defined as before by considering only the two lowest levels. Although the large $E_{\rm J} / E_{\rm C}$ ratio makes the phase qubit insensitive to charge noise, there is not, unlike for the charge and flux qubits, any symmetry point where the phase qubit is particularly well protected from noise sources. 


\section{Further Josephson-junction qubits}
\label{sec:FurtherQubits}

To scale up Josephson-junction qubits for large-scale quantum computation, it is essential that the quantum coherence of the qubits can be maintained for as long as possible. Through the years, many refinements of the three basic circuit designs reviewed in the previous section have been proposed and tested, mostly with the aim of improving coherence, but also for purposes like increasing connectivity or the tunability of parameters. In this section, we first explain the workings of a currently popular design, the transmon qubit~\cite{Koch2007}, and then give an overview of other updates to the basic qubit designs. Note that there also exist proposals for other Josephson-junction-qubit designs that do not build directly on the three basic circuits; examples include so-called phase-slip qubits, Andreev-level qubits, and d-wave qubits~\cite{Zagoskin2007}. 


\subsection{The transmon qubit}
\label{sec:Transmon}

\begin{figure}
\centering
\includegraphics[width=\linewidth]{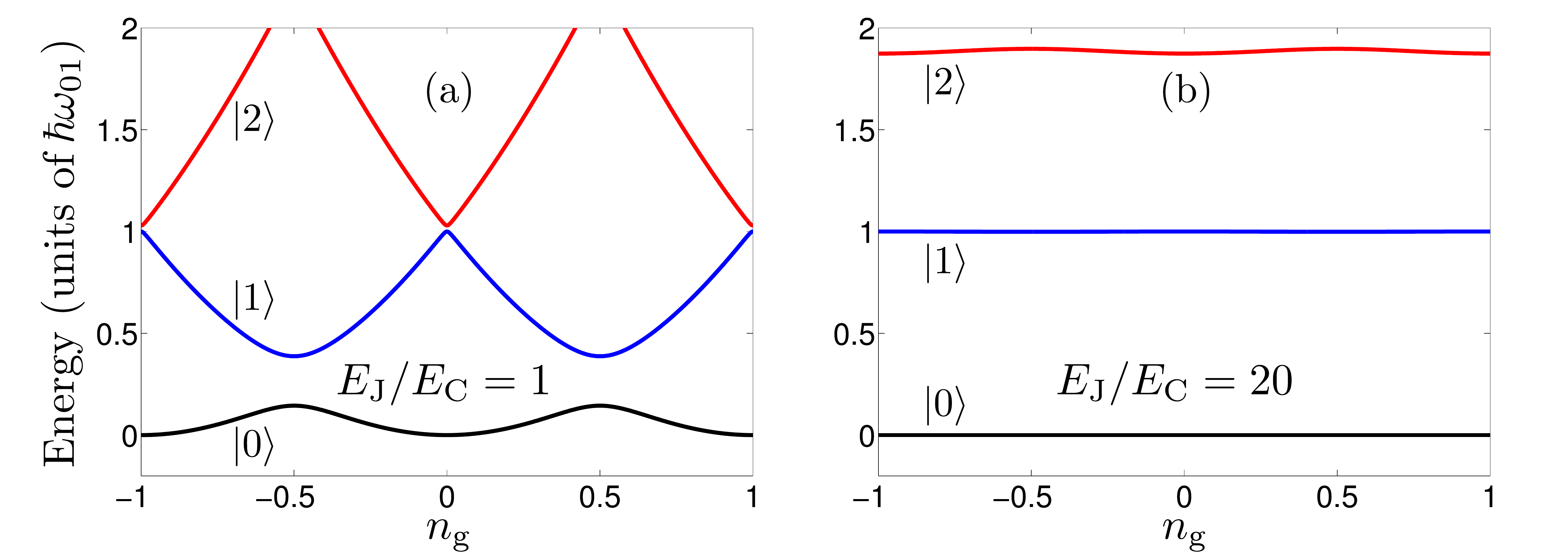}
\caption{Energy levels of a CPB for different $E_{\rm J} / E_{\rm C}$ ratios.
(a) $E_{\rm J} / E_{\rm C} = 1$. This is the charge-qubit regime, where, as explained in \secref{sec:ChargeQubit}, a good qubit is formed when $n_{\rm g} \approx \pm 0.5$. At these points, $\omega_{01}$ is nowhere close to $\omega_{12}$ and the transition frequencies are not so sensitive to fluctuations in $n_{\rm g}$.
(b) $E_{\rm J} / E_{\rm C} = 20$. This is the transmon-qubit regime, where the energy levels are insensitive to fluctuations in $n_{\rm g}$ no matter what the value of $n_{\rm g}$ is. The anharmonicity of the energy-level spacing is less than for the charge qubit, but still enough to make a good qubit.
\label{fig:TransmonEnergyLevels}}
\end{figure}

The \textit{transmon qubit} (the name was originally an abbreviation of the unwieldy ``transmission-line shunted plasma oscillation qubit'') is formed by adding another capacitance $C_{\rm B}$, in parallel with the Josephson junction, to the charge-qubit circuit in \figref{fig:ThreeBasicQubits}(a)~\cite{Koch2007}. This is similar to an earlier proposal that modifies a flux qubit in the same way~\cite{You2007}.  Adding the extra capacitance decreases the charging energy $E_{\rm C}$ in the circuit. By changing the $E_{\rm J} / E_{\rm C}$ ratio from $E_{\rm J} / E_{\rm C} \approx 10^{-1}$ to $E_{\rm J} / E_{\rm C} \approx 10^{2}$, the charge-qubit circuit goes from having a well-defined $n$ to having a well-defined $\phi$. However, the resulting energy levels are largely \textit{insensitive} to fluctuations in $n_{\rm g}$, as shown in \figref{fig:TransmonEnergyLevels}.

The price one pays for this protection from charge noise is a \textit{decrease in the anharmonicity} of the circuit. In the limit $E_{\rm J} \gg E_{\rm C}$, perturbation theory in the small variable $E_{\rm C} / E_{\rm J}$ gives that the energy levels $E_m$ of the circuit are well approximated by~\cite{Koch2007}
\be
E_m = - E_{\rm J} + \sqrt{8 E_{\rm J} E_{\rm C}} \mleft( m + \frac{1}{2} \mright) - \frac{E_{\rm C}}{12} \mleft( 6 m^2 + 6 m + 3 \mright).
\ee
From this, we obtain the qubit transition frequency
\be
\omega_{01} = \mleft( \sqrt{8 E_{\rm J} E_{\rm C}} - E_{\rm C} \mright) / \hbar
\ee
and the anharmonicity
\be
\omega_{12} - \omega_{01} = - E_{\rm C} / \hbar.
\ee
However, the trade is a favorable one. A detailed analysis using perturbation theory shows that the decrease in sensitivity to charge noise is \textit{exponential} in $\sqrt{E_{\rm J} / E_{\rm C}}$, while the anharmonicity only decreases \textit{linearly} in $\sqrt{E_{\rm J} / E_{\rm C}}$ when scaled by $\omega_{01}$. Recall that $E_{\rm J} / E_{\rm C}$ can be tuned by an external magnetic flux if the Josephson junction is replaced by a SQUID [\eqref{eq:TunableEJ}].


\subsection{Other qubit refinements}
\label{sec:OtherRefinements}

\begin{figure}
\centering
\includegraphics[width=\linewidth]{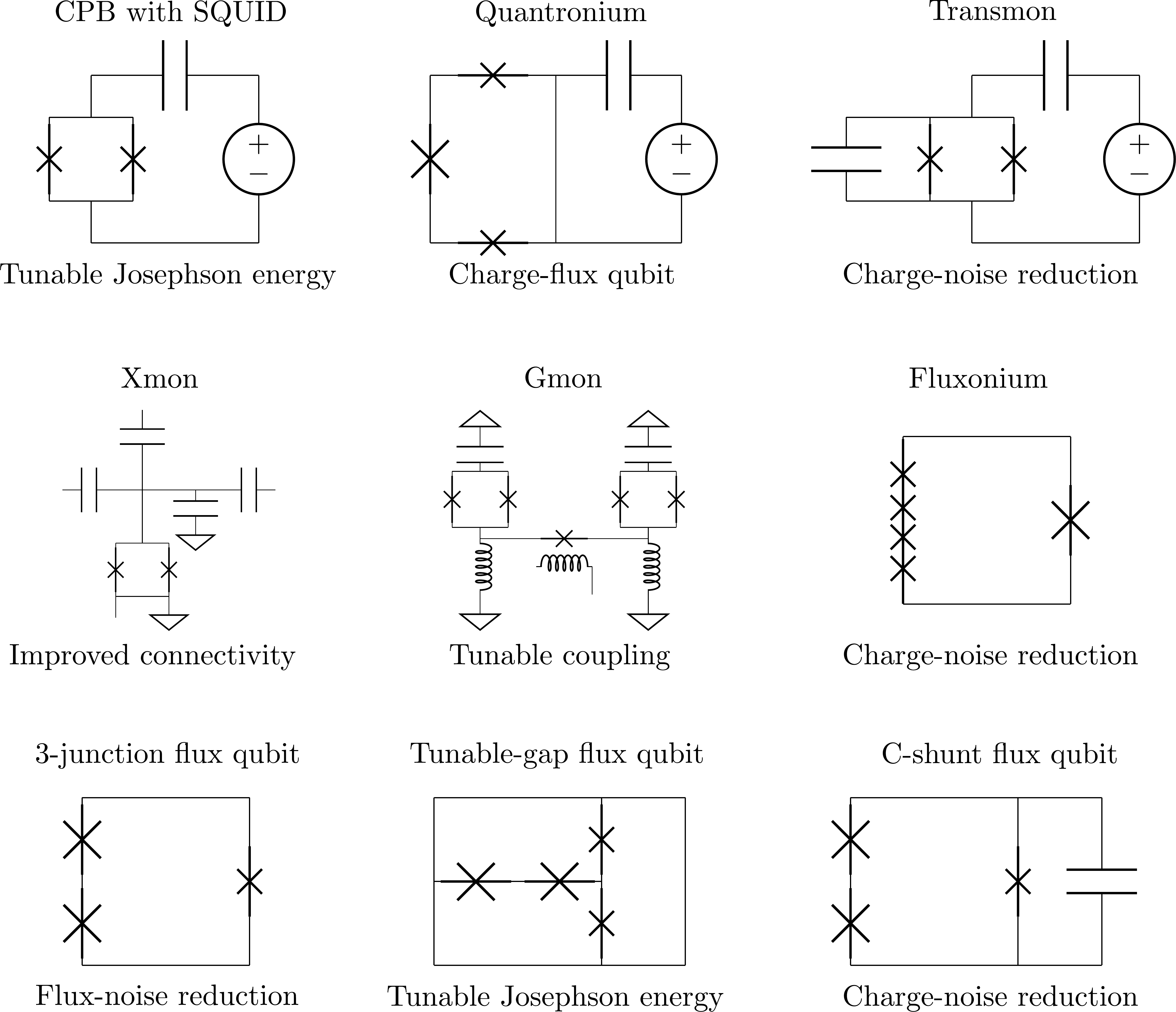}
\caption{A chart of various extensions and refinements of the three basic Josephson-junction qubits. For each circuit design, the name is written above the circuit and the main improvement in performance is listed below the circuit. More details are given in the text.
\label{fig:QubitRefinements}}
\end{figure}

An overview of extensions of the three basic qubit designs of \secref{sec:ThreeQubits} is presented in \figref{fig:QubitRefinements}. We have already mentioned in preceding sections that replacing a Josephson junction with a SQUID makes it possible to tune $E_{\rm J}$. In the top left corner of \figref{fig:QubitRefinements}, such a replacement is shown for a charge qubit (Cooper-pair box) and in the center of the bottom row of \figref{fig:QubitRefinements}, the same idea is applied to a flux qubit~\cite{Paauw2009}. We also note that recently there have been experiments with transmon qubits where the Josephson junction is formed by two superconductors connected through a semiconductor nanowire~\cite{Larsen2015, DeLange2015}. This has been called a \textit{gatemon} circuit, since here $E_{\rm J}$ can be tuned by a gate voltage applied to the nanowire.

The most important motivation for improving qubit design has been to extend qubit \textit{coherence time}, i.e., the time that the quantum coherence of the qubit is preserved before being lost due to noise from the surroundings. The first charge and flux qubits only had coherence times of a few nanoseconds. Remember that most Josephson-junction qubits have transition frequencies $\omega_{01}$ in the range \unit[1-10]{GHz}. One of the first improvements on the aforementioned short coherence times was to combine features of these two designs to make the \textit{quantronium}~\cite{Vion2002} qubit, shown in the center of the upper row in \figref{fig:QubitRefinements}. Operating in an intermediate regime where $E_{\rm J} / E_{\rm C} \approx 1$, this circuit boosted coherence times to about \unit[500]{ns}.

Another strategy for reducing environmental noise that we already have mentioned is the addition of a capacitance to the charge qubit, forming the transmon circuit shown in the top right of \figref{fig:QubitRefinements}. This increases $E_{\rm J} / E_{\rm C}$ and makes the qubit less sensitive to charge noise. Transmon qubits have reached coherence times on the order of \unit[100]{$\mu$s}~\cite{Riste2013, Jin2015}.

The same trick of adding a shunt capacitance to decrease sensitivity to charge noise has also been applied to the flux qubit~\cite{You2007, Yan2016}, as shown in the bottom right of \figref{fig:QubitRefinements}. This has also resulted in greatly improved coherence times, although, just as in the case of the transmon, there is a price to be paid in the form of a decrease in anharmonicity of the circuit. The increased need for protection from charge noise in the flux qubit arose due to the introduction of the 3-junction flux qubit (bottom left in \figref{fig:QubitRefinements}), which made it possible to reduce loop size and thus reduced flux noise, but made the circuit more sensitive to charge noise instead. We note that there also is a design with a shunt capacitance added to a phase qubit~\cite{Steffen2006}.

A further development of the flux-qubit design is the \textit{fluxonium}~\cite{Manucharyan2009} qubit, shown center right in \figref{fig:QubitRefinements}. In this design, one Josephson junction is shunted by an array of Josephson junctions, which suppress charge noise by having large capacitances, but also help achieving a high anharmonicity by providing a large inductance. An experiment with a fluxonium qubit~\cite{Pop2014} is at the time of writing the only that ever demonstrated a Josephson-junction qubit being protected from energy relaxation for more than one millisecond.

There are also design developments building on the noise-resistant transmon qubit with an eye to scaling up to circuits containing many coupled qubits. One such design is the \textit{xmon}~\cite{Barends2013} qubit shown center left in \figref{fig:QubitRefinements}. By making the superconducting island cross-shaped, this version of the transmon can be capacitively coupled to multiple other qubits and/or control lines. For coupling two transmon qubits directly, there is also the \textit{gmon}~\cite{Chen2014} circuit shown in the center of \figref{fig:QubitRefinements}. The coupling between the two qubits can be tuned inductively during the experiment.


\section{Quantum computing with Josephson-junction qubits}
\label{sec:QC}

As described in \secref{sec:Intro}, the main motivation for the development of Josephson-junction qubits has been their potential application as building blocks for a quantum computer. This is the reason why today companies like IBM and Google have large teams of researchers trying to make a significant number of Josephson-junction qubits work well together. At the time of writing, these research groups are approaching system sizes of almost 100 qubits. In this section, we provide a more detailed overview of why Josephson-junction qubits are seen as one of the most promising platforms for quantum computation. We also discuss related topics like other approaches to quantum computation, quantum simulation, and quantum error correction. For more in-depth reviews of this topic, see Refs.~\cite{Wendin2017, Gu2017, Devoret2013}.


\subsection{Fulfilling the DiVincenzo criteria}

When judging how suitable a physical system is for building a universal digital quantum computer (i.e., a qubit-based computer, using gates, that in theory can be programmed to do anything given enough time and resources), the gold standard is the \textit{DiVincenzo criteria}~\cite{DiVincenzo2000}. These are five conditions that need to be fulfilled by Josephson-junction qubits, trapped ions, or any other prospective gate-based quantum-computing architecture. The criteria are
\begin{enumerate}[(DV1)]
\item Qubits: it must be possible to fabricate multiple qubits.
\item Initialization: it must be possible to initialize these qubits to a simple, known state, e.g., $\ket{000 \ldots 00}$.
\item Gates: it must be possible to perform both single- and two-qubit gates on the qubits with high fidelity. Single-qubit gates are rotations on the Bloch sphere; two-qubit gates are quantum versions of classical two-bit gates like XOR or controlled-NOT. The set of available gates must be \textit{universal}, i.e., they must together enable any conceivable program to be implemented on the quantum computer.
\item Readout: it must be possible to measure the states of the qubits.
\item Coherence: the coherence times of the qubits must be long enough to allow a large number of gates to be performed in sequence before a significant loss of quantum coherence occurs.
\end{enumerate}
To these five criteria, one can also add that it is desirable to 
\begin{enumerate}[(i)]
\item Have an interface that can transmit quantum information from the qubits in the computer to qubits in a memory or to qubits used for long-distance communication.
\item Be able to communicate quantum information across long distances.
\end{enumerate}
To realize (i), a promising route is \textit{hybrid quantum systems}, where superconducting circuits couple to some other type of system, which may not be suitable for quantum computation itself, but has excellent coherence times instead~\cite{Buluta2011, Xiang2013}. This other system could even be one of the two-level systems that occur naturally in a Josephson junction~\cite{Zagoskin2006, Neeley2008}. For (ii), optical photons are ideal information carriers~\cite{Kimble2008}. There is currently much effort being devoted to designing devices that can convert quantum information from the microwave frequencies of Josephson-junction qubits to optical frequencies. Since these energy scales differ by roughly five orders of magnitude, it is very hard to achieve good conversion efficiency~\cite{Wendin2017}.


\subsubsection{Qubits}

Regarding (DV1), we have already shown in this chapter that there is a multitude of different Josephson-junction qubits available. Since these circuits are easy to fabricate on a chip, they can be scaled up to systems with many qubits. However, it remains an outstanding engineering challenge to scale up the connections to control electronics needed to manipulate and read out the many qubits in such a large system.


\subsubsection{Initialization}

When it comes to (DV2), several methods are available to initialize Josephson-junction qubits in a known state (usually the ground state $\ket{0}$). One method is to simply measure the qubit, projecting it into $\ket{0}$ (or flipping it from $\ket{1}$ to $\ket{0}$ through a simple rotation if the measurement result is $\ket{1}$)~\cite{Riste2012, Johnson2012}. One can also control the environment of the qubit to induce relaxation to $\ket{0}$~\cite{Reed2010} or use a driven setup where the steady state has the qubit in $\ket{0}$~\cite{Geerlings2013}.


\subsubsection{Gates}
\label{sec:Gates}

For (DV3), we first note that single-qubit rotations together with any ``non-trivial'' two-qubit gate constitutes a universal gate set. One example of such a two-qubit gate is the controlled-NOT (CNOT) gate, which flips qubit 2 if qubit 1 (the controlling qubit) is in its excited state:
\be
\ket{00}, \ket{01}, \ket{10}, \ket{11} \xrightarrow{\text{CNOT}} \ket{00}, \ket{01}, \ket{11}, \ket{10}.
\label{eq:CNOT}
\ee
Another useful two-qubit gate is the controlled-phase (CPHASE) gate, which adds a phase factor $e^{i \varphi}$ to the state $\ket{11}$ and leaves all other states unchanged:
\be
\ket{00}, \ket{01}, \ket{10}, \ket{11} \xrightarrow{\text{CPHASE}} \ket{00}, \ket{01}, \ket{10}, e^{i \varphi} \ket{11}.
\ee
The special case $\varphi = \pi$ is known as the controlled-Z (CZ) gate. A third two-qubit gate is the iSWAP gate, which leaves $\ket{00}$ and $\ket{11}$ unchanged, but swaps the states $\ket{01}$ and $\ket{10}$ into each other, adding a factor $i$ in front of them:
\be
\ket{00}, \ket{01}, \ket{10}, \ket{11} \xrightarrow{\text{iSWAP}} \ket{00}, i\ket{10}, i\ket{01}, \ket{11}.
\ee

All the above two-qubit gates have been implemented with Josephson-junction qubits. At the time of writing, state of the art for single-qubit gates is fidelities above 99.9\%~\cite{Barends2014, Sheldon2016} and above 99\% for two-qubit gates~\cite{Barends2014, Sheldon2016a}. In the work of Ref.~\cite{Barends2014}, the two-qubit gate used was the CPHASE gate. It was realized with gmon qubits (see \secref{sec:OtherRefinements}) and utilized the second excited state of these qubits, achieving a phase shift of the state $\ket{11}$ only by bringing it into resonance with the state $\ket{02}$ for a short time. In Ref.~\cite{Sheldon2016a}, the two-qubit gate was instead a CNOT one. Unlike the previous example, this gate used a scheme called cross-resonance, which does not require tuning any qubit frequency. Instead, the gate is implemented between two qubits with different transition frequencies. Both qubits are driven at their respective transition frequencies, but one of them is also driven at the transition frequency of the other. Since no qubit frequency need to be tuned, the gate can be implemented with transmons containing only single Josephson junctions. The absence of a SQUID in the transmon makes that qubit less sensitive to flux noise, so a single-junction transmon generally has longer coherence times.

We also mention that, similar to the CPHASE implementation described above, the three-qubit Toffoli gate has been implemented by taking advantage of higher energy levels in transmon qubits~\cite{Fedorov2012, Reed2012}. The Toffoli gate can be seen as a CNOT gate with two control qubits. It can also form the basis for universal quantum computation. For a more detailed review of gates in superconducting circuits, see Ref.~\cite{Wendin2017}.


\subsubsection{Readout}

There are many ways to measure the states of Josephson-junction qubits (for a more detailed overview, see Refs.~\cite{Gu2017, Siddiqi2011}). For measurements on the three basic Josephson-junction-qubit designs in \secref{sec:ThreeQubits}, there are observables in the circuits that can be accessed directly. In charge qubits, one can measure the charge on the superconducting island, e.g., using a single-electron transistor~\cite{Nakamura1999}. Since the charge qubit states are in the charge eigenbasis, such a measurement directly gives information about the qubit state. In flux qubits, a nearby SQUID can be used to detect the direction of circulation for the persistent current in the flux qubit loop, which determines the qubit state~\cite{VanderWal2000}. For the phase qubit, the bias current is tuned such that, in the tilted-washboard potential, the probability for tunneling out of the potential well is much greater if the qubit is in state $\ket{1}$. Tunneling switches the voltage state of the Josephson junction, which is easy to detect~\cite{Martinis1985}.

However, the measurements above are not \textit{quantum nondemolition} (QND), i.e., they do not preserve the state that the measurement projects the qubit into. In modern setups, Josephson-junction qubits are usually read out in a QND way by having them coupled to a resonator with frequency $\omega_{\rm r}$. When the qubit transition frequency $\omega_{01}$ is far detuned from $\omega_{\rm r}$, i.e., when
\be
\abs{\omega_{\rm r} - \omega_{01}} \gg g,
\ee
where $g$ is the strength of the coupling between the qubit and the resonator, the system is said to be in the dispersive regime. In this regime, no excitations are exchanged between the qubit and the resonator due to the mismatch in frequencies. However, the coupling gives rise to a shift of $\omega_{\rm r}$ that depends on the qubit state~\cite{Blais2004}. Thus, by probing the cavity, the qubit state can be inferred indirectly.

Note that a measurement like the dispersive one does not need to be projective. If only a weak signal is used to probe the cavity, information about the qubit state is acquired gradually, not all at once. Such a \textit{weak measurement} (not to be confused with weak-value measurements) can sometimes even be reversed~\cite{Katz2008}. To distinguish the weak signal, amplifiers are needed. Since amplifiers like high-electron mobility transistors (HEMTs) add too much noise at the low temperatures where superconducting circuits operate, much effort has been devoted to develop on-chip cryogenic amplifiers based on Josephson junctions~\cite{Clerk2010, Nation2012, Roy2016}. If the noise is too large, it is necessary to average over many experimental runs to infer the qubit state, but with the aid of Josephson-junction-based amplifiers, Josephson-junction qubits can be read out in a single experimental run (``single-shot measurement'')~\cite{Mallet2010}.


\subsubsection{Coherence}

When discussing coherence times, note that there are \textit{three} different times that are all often quoted. There is $T_1$, the timescale for energy relaxation, i.e., the time after which decay to $\ket{0}$, induced by the qubit environment, has changed the probability to find a qubit initialized in state $\ket{1}$ from 1 to $1/e$. There is also $T_2$, the timescale on which the phase coherence between the qubit states is preserved. If the only decoherence process is energy relaxation, $T_2 = 2 T_1$. If there is some other process that causes pure dephasing, characterized by a timescale $T_\varphi$, the decoherence times are related via
\be
\frac{1}{T_2} = \frac{1}{2 T_1} + \frac{1}{T_\varphi}.
\ee

We already showed in \secref{sec:OtherRefinements} how refinements of the design for Josephson-junction qubits have increased coherence times dramatically, from a few nanoseconds to hundreds of microseconds or even a millisecond. For a more detailed overview of this development, see Refs.~\cite{Gu2017, Oliver2013}. Since gate operations typically take on the order of \unit[10-100]{ns}, it is now feasible to talk about performing many gates, as well as initialization and readout, while quantum coherence is preserved.


\subsubsection{Tunable coupling}

Although not explicitly part of the DiVincenzo criteria, the ability to control, in situ, the coupling between Josephson-junction qubits (and possibly to other circuits element) is highly desirable for scaling up to realize a quantum computer. In particular, this ability is important for implementing many types of gates (see \secref{sec:Gates}).

Josephson-junction qubits can be connected either directly, capacitively or inductively, or via some intermediate coupling element connected to both qubits, e.g., an $LC$ resonator or another Josephson-junction qubit. To turn qubit-qubit couplings on and off, one method is to tune (e.g., by adjusting the flux through SQUID loops in the qubits; see \secref{sec:OtherRefinements}), the transition frequencies of the two qubits far from resonance with each other~\cite{Blais2003, Berkley2003, McDermott2005, Majer2007, Sillanpaa2007}. This method has been used in some recent two-qubit-gate implementations~\cite{Corcoles2015, Barends2016}. However, this method has some drawbacks. One drawback is that in larger circuits, where more qubits are coupled, it may be hard to find frequency values such that all neighboring qubits are detuned from each other. Furthermore, the frequency tuning should be adiabatic to preserve the qubit states. Finally, frequency-tunable qubits are generally less coherent than fixed-frequency ones: partly because tuning the frequency can take a qubit away from its optimal working point, partly because the tunability mechanism can be affected by noise that results in dephasing.

The main alternative to tuning the qubits themselves is to instead tune the element connecting them~\cite{Makhlin1999, You2002, Rigetti2005, Liu2006, Grajcar2006, Ashhab2008}. Such schemes have also seen widespread experimental implementation~\cite{Hime2006, Niskanen2007, Bialczak2011, Chen2014, McKay2016}. However, also this method has drawbacks. The tunable coupling elements take up space on the chip and provide new channels through which noise can affect the qubits. Thus, the search continues for new methods for tunable coupling~\cite{Wu2018}.


\subsubsection{Summary}

In summary, all five DiVincenzo criteria have been fulfilled, at least to a reasonable degree, in experiments with Josephson-junction qubits. This is very promising for superconducting quantum computation, but several issues, both fundamental and engineering ones, remain to be solved before a large-scale universal quantum computer based on Josephson junctions becomes reality.


\subsection{Adiabatic quantum computing and quantum annealing}

The approach of the previous section, universal gate-based quantum computation, is not the only way to perform calculations with Josephson-junction (or other) qubits. One alternative is \textit{adiabatic quantum computing}~\cite{Farhi2001} (AQC; for a recent review, see Ref.~\cite{Albash2018}). The strategy employed in AQC is to set up a system of qubits governed by a simple Hamiltonian, prepare this system in its ground state, and then adiabatically change the Hamiltonian to a more complex one, whose ground state contains the solution of the problem the computation is meant to solve. Here, ``adiabatically'' means that the parameters of the system are changed slowly enough that the system remains in its ground state throughout the evolution.

The idea of AQC is appealing; AQC is theoretically equivalent~\cite{Aharonov2004} to universal quantum computing, but it can be more robust against noise than the gate-based version. However, the computational speed that has to be sacrificed in order to ensure adiabatic time evolution means that it is not clear whether AQC actually can provide any speed-up compared to classical computation. For this reason, there have not been many implementations of AQC with Josephson-junction qubits. At the time of writing, the most advanced example is an experiment~\cite{Barends2016} where nine gmon qubits used a combination of gate-based and adiabatic quantum computing to solve the 1D Ising model and some other Hamiltonians. 

Another computational method is \textit{quantum annealing} (QA)~\cite{Das2008}, which builds on the classical computation method known as simulated annealing. In simulated annealing, artificial thermal fluctuations aid the search for the solution of an optimization problem by helping the search overcome energy barriers of local minima. In QA, a system is initialized in some state (not necessarily the ground state) at non-zero temperature and then evolves into the ground state of a Hamiltonian which encodes the problem to be solved. For certain potential-energy landscapes with tall and high barriers, the effect of quantum tunneling can provide a boost to the search that is absent in simulated annealing. 

Quantum annealing has seen more experimental investigations with Josephson-junction qubits than AQC. These experiments are the superconducting quantum computations that use the most qubits to date; there are several examples with hundreds of qubits, e.g., Refs.~\cite{Boixo2014, Pudenz2014}. There are even larger circuits, with up to 2048 qubits, manufactured by the company D-Wave. However, it is still unclear whether these systems, which suffer from issues with decoherence and connectivity, can actually provide a significant speed-up compared to classical computational methods~\cite{Ronnow2014, Zagoskin2014, Heim2015, Denchev2016}.


\subsection{Quantum simulation}

As noted in \secref{sec:WhatIsAQubit}, the original motivation~\cite{Feynman1982} for trying to make qubits was not gate-based quantum computation, but the frustrating difficulty of using classical bits to \textit{simulate} the behavior of quantum systems. Such \textit{quantum simulation}, reviewed in more detail in Refs.~\cite{Buluta2009, Georgescu2014}, is a more easily achievable and nearer-term goal than full-fledged universal quantum computation. Superconducting circuits with Josephson-junction qubits are well suited to quantum-simulation applications, since they can be arranged in many setups and can have parameters, like transition frequencies and coupling strengths, tuned during an experiment. Note that there are two approaches to quantum simulation:
\begin{enumerate}[(i)]
\item Analog quantum simulation, where the qubits are arranged to directly emulate the system of interest.
\item Digital quantum simulation, where algorithms are implemented on a gate-based quantum computer to simulate the system.
\end{enumerate}

Already a \textit{single} Josephson-junction qubit turns out to be quite powerful for quantum simulation. Since Josephson-junction qubits also have higher excited states, they can be used to emulate the behavior of large spins. When using $d > 2$ levels in the circuit, one can speak of Josephson-junction \textit{qudits} instead. For example, a Josephson-junction \textit{quintit} ($d = 5$) has been used to simulate the dynamics of spins with sizes up to $S = 3/2$~\cite{Neeley2009, Nori2009}. We note that single Josephson-junction qubits also have been used for quantum simulation of topological quantum phenomena~\cite{Schroer2014, Roushan2014}.

Scaling up to more than one Josephson-junction qubit, a triangular loop of three coupled such qubits has been used to simulate various properties of interacting photons, including synthetic magnetic fields and strong photon-photon interactions~\cite{Roushan2017}. The fractional statistics of anyons has been simulated in a superconducting circuit with four qubits and one resonator~\cite{You2010, Zhong2016}. Another phenomenon from condensed-matter physics that has been demonstrated is weak localization~\cite{Chen2014a}. 

We also note that there is great interest in quantum simulation based on large lattices of superconducting resonators coupled to Josephson-junction qubits~\cite{Houck2012, Noh2017}. Recent experimental examples include a 49-site Kagome lattice~\cite{Underwood2016} and a 72-site 1D lattice that was used to demonstrate a dissipative phase transition~\cite{Fitzpatrick2017}.

More interesting for practical applications are perhaps simulations of \textit{molecules} that allow for calculation of energies in such systems. Recently, a few Josephson-junction qubits have been used to calculate ground-state energies for hydrogen~\cite{OMalley2016} and $\text{BeH}_2$~\cite{Kandala2017} molecules, but it is unclear if the approach in these experiments scales well when moving to larger systems. An example of such a large system, which constitutes an enticing goal of \textit{quantum chemistry}, is the enzyme nitrogenase. Today, nitrogen for fertilizer is extracted through the so-called Haber-Bosch process, which is energy-demanding; more than 1\% of the world's total energy consumption is estimated to power this process. Nitrogenase, which is produced by certain bacteria, can perform the process much more efficiently, at room temperature. Despite many efforts, the mechanism used by this enzyme is not yet known, but it is estimated that a future quantum computer or simulator could help provide the missing information~\cite{Reiher2017}. However, this calculation still seems to require on the order of a million qubits, with improved gates and coherence times, which is a daunting task.



\subsection{Quantum error correction}

Although single- and two-qubit gates can be performed with high precision (see \secref{sec:Gates}) and Josephson-junction qubits have long coherence times (see \secref{sec:OtherRefinements}), the error rate in qubits is still much higher than for modern classical bits. For truly fault-tolerant quantum computation, some form of \textit{quantum error correction} (QEC)~\cite{Nielsen2000, Devitt2013, Terhal2015} is necessary.

For classical bits, it is easy to design an error-correction scheme: simply make two copies of the bit that carries the information you wish to protect. When it is time to read out the information in the bit, measure all three bits and let a majority vote among them decide the result. If the error probability for a single bit is $p \ll 1$, then the probability that the majority vote gives the wrong result is proportional to $p^2$, which is a great improvement. However, this scheme cannot be directly applied to qubits, for several reasons:
\begin{itemize}
\item It is impossible to clone arbitrary quantum states~\cite{Wootters1982, Dieks1982}. 
\item Measuring a qubit will project it into one of its eigenstates, destroying any superposition state. 
\item The only error a classical bit can suffer is a bit flip, but an error on a quantum bit can be any rotation on the Bloch sphere.
\end{itemize}
%

\begin{figure}
\centering
\includegraphics[width=\linewidth]{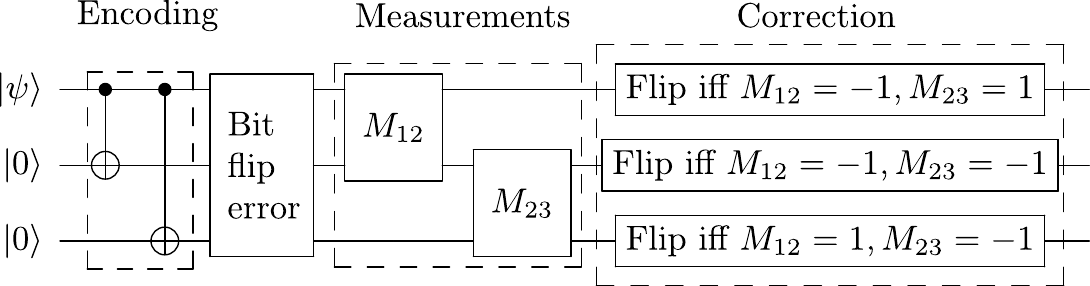}
\caption{The 3-qubit QEC for bit-flip errors. The state $\ket{\psi}$ is entangled with two other qubits through CNOT gates. Parity measurements on pairs of qubits are used to detect bit-flip errors. Depending on the results of these measurements, qubits are flipped to reset the system state to what it was before the error occurred.
\label{fig:3QubitQEC}}
\end{figure}

Fortunately, it was worked out in the 1990s how to overcome these obstacles~\cite{Shor1995, Steane1996, Knill1997}. As an example, here we explain the three-qubit code for correcting bit-flip errors. The scheme is shown in \figref{fig:3QubitQEC}. We have a qubit in the general state $\ket{\psi} = \alpha \ket{0} + \beta \ket{1}$ that we wish to protect from bit flips. By performing CNOT gates [\eqref{eq:CNOT}] with this qubit as the control and two qubits in states $\ket{0}$ as targets, the three-qubit state becomes
\be
\ket{\psi_3} = \alpha\ket{000} + \beta\ket{111}.
\label{eq:Psi3}
\ee
Note that this \textit{entangled} state is different from the \textit{separable} state
\be
\ket{\psi_{3, \rm sep}} = \mleft( \alpha \ket{0} + \beta \ket{1} \mright) \mleft( \alpha \ket{0} + \beta \ket{1} \mright) \mleft( \alpha \ket{0} + \beta \ket{1} \mright)
\ee
that could be created if quantum cloning was possible.

We now assume that the third qubit suffers a bit-flip error. The resulting system state is then
\be
\ket{\psi_{3, \rm err}} = \alpha \ket{001} + \beta \ket{110}.
\label{EqPsi3err}
\ee
To detect this error without destroying the superposition, we use \textit{parity measurements}, i.e., multi-qubit measurements which only reveal whether an odd or even number of the qubits are in the same state, nothing more. Performing a parity measurement on qubits 1 and 2, we see that they are in the same state: $\ket{00}$, $\ket{11}$, or some superposition of the two. Measuring the parity of qubits 2 and 3, we see that one of them has been flipped. Assuming that the probability for more than one bit flip having occurred is negligible, we can thus conclude that qubit 3 was flipped and apply a rotation to this qubit to reset the system to $\ket{\psi_3}$.

This example demonstrates how to overcome the first two hurdles of QEC listed above. For the third, the fact that qubit errors can be arbitrary rotations on the Bloch sphere, it turns out that it is enough to combine schemes detecting flips along various axes to also correct for small rotation errors.

The three-qubit bit-flip correction code has been demonstrated with Josephson-junction qubits~\cite{Reed2012}. A later experiment extended this principle of using parity measurements on pairs of qubits to detect errors to a 1D array of nine Josephson-junction qubits~\cite{Kelly2015}. However, a more promising architecture for truly large-scale error correction in superconducting circuits is 2D \textit{surface codes}~\cite{Bravyi1998, Fowler2012}. In these codes, qubits are positioned with nearest-neighbor couplings on a square lattice. Half of the qubits are used for computation; the other half are used to measure the four-qubit parities of their four neighbors through consecutive CNOT gates. Such four-qubit parity measurements have been demonstrated in an experiment using transmon qubits~\cite{Takita2016}. Provided that gates reach high enough fidelity (Josephson-junction qubits are at this threshold~\cite{Barends2014}), scaling up the size of the 2D lattice will make it possible to preserve one qubit of information, often referred to as a logical qubit, for a very long time.

We note that Josephson-junction qubits often are connected to resonators whose coherence times can exceed those of the qubits. For this reason, another approach to QEC in superconducting circuits is to encode the quantum information in the photonic states of these resonators. These error-correction codes are known as bosonic codes~\cite{Albert2018}. Recently, an experimental implementation~\cite{Ofek2016} of such a code in superconducting circuits reached ``break-even'' for QEC, i.e., the coherence time of the logical qubit exceeded the coherence times of all parts making up the system.


\section{Quantum optics and atomic physics with Josephson-junction qubits}
\label{sec:QO}

Josephson-junction qubits are not only a promising platform for quantum computation; they are also an excellent tool for exploring fundamental questions in quantum optics and atomic physics. The advantages of Josephson-junction qubits for QIP, listed at the beginning of \secref{sec:WhyJJQubits}, also facilitate exploration of light-matter interaction at the quantum level with these systems instead of natural atoms and optical photons. Furthermore, Josephson-junction qubits have opened the door to new regimes of quantum optics. In the following, we illustrate these exciting developments with a few examples. For a more detailed overview of quantum optics and atomic physics in superconducting circuits, see Refs.~\cite{You2011, Gu2017}.


\subsection{New prospects for textbook quantum optics}

The typical setup for studying interaction between light and matter in quantum optics is to have one or more natural atoms confined in a small cavity. The atoms interact with the photons in the quantized electromagnetic modes of the cavity. This is known as \textit{cavity quantum electrodynamics} (QED)~\cite{Haroche2013}.

\begin{figure}
\centering
\includegraphics[width=0.75\linewidth]{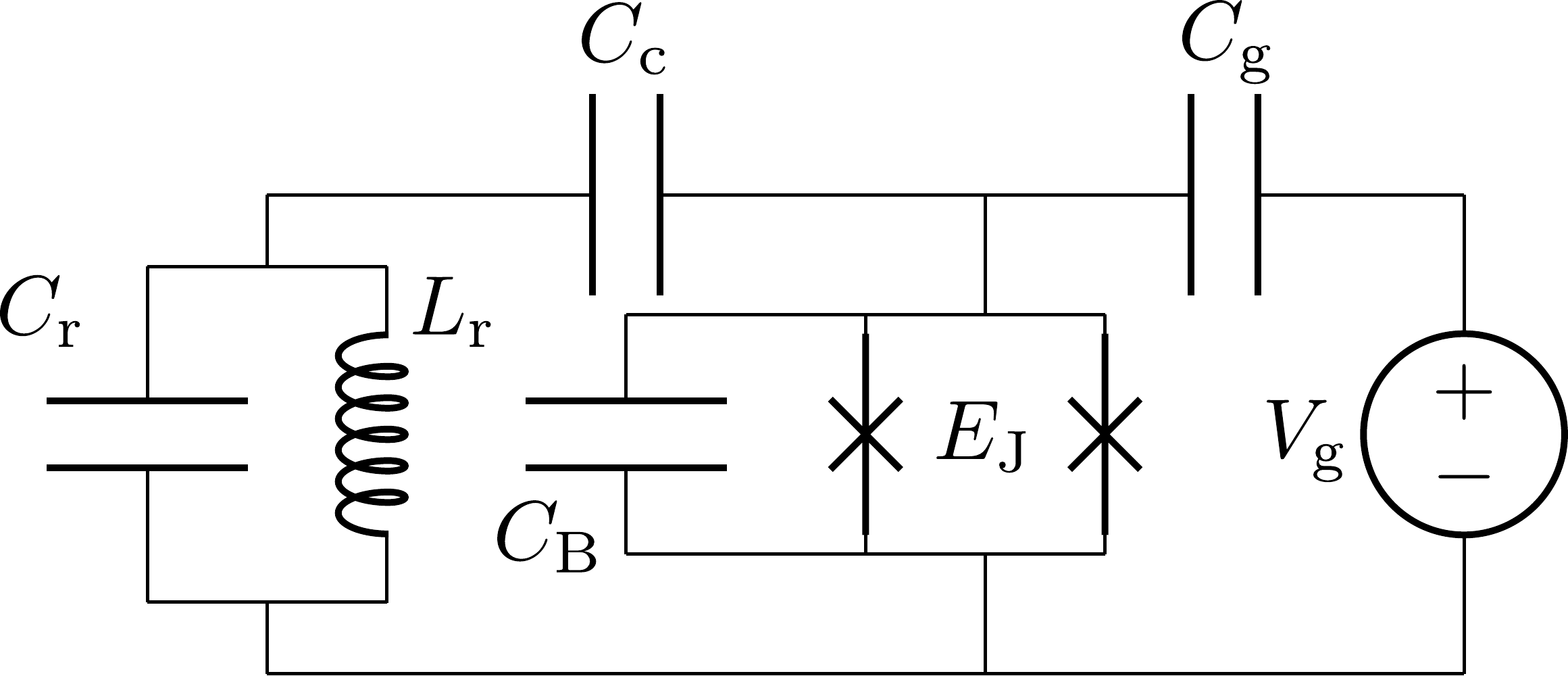}
\caption{A circuit-QED setup. The transmon qubit (right; see \secref{sec:Transmon}) is connected via a coupling capacitance $C_{\rm c}$ to the resonator formed by $L_{\rm r}$ and $C_{\rm r}$.
\label{fig:CircuitQED}}
\end{figure}

A similar setup can be realized with Josephson-junction qubits~\cite{You2003, Blais2004}. In this case, the optical cavity is replaced by an $LC$ or transmission-line resonator (or a microwave cavity). This is known as \textit{circuit QED}. A typical such circuit, featuring a capacitive coupling between a transmon qubit and an $LC$ resonator, is shown in \figref{fig:CircuitQED}.

In both cavity and circuit QED, the system can usually be described by the Jaynes--Cummings Hamiltonian~\cite{Jaynes1963, Gerry2005}
\be
H_{\rm JC} = \hbar \omega_{\rm r} a^\dag a + \hbar \omega_{01} \sz + \hbar g \mleft( a^\dag \sm + a \sp \mright).
\label{eq:JC}
\ee
Here, $\omega_{\rm r}$ is the resonance frequency of the photonic mode, $a^\dag$ ($a$) is the creation (annihilation) operator for photons, $\omega_{01}$ is the transition frequency of the qubit, $\sp$ ($\sm$) is the raising (lowering) operator for the qubit, and $g$ is the light-matter coupling strength. Note that the coupling term either converts a photon into a qubit excitation or vice versa.

The field of circuit QED took off in 2004 when an experiment~\cite{Wallraff2004} demonstrated \textit{strong coupling} between a charge qubit and a transmission-line resonator. Strong coupling is defined as $g$ exceeding the decoherence rates of both the qubits and the resonator. Reaching strong coupling is important because it means excitations can be exchanged between the qubit and the resonator before the quantum coherence of the system is lost. In recent circuit-QED experiments, $g$ is often several orders of magnitude larger than the decoherence rates. This is very hard to achieve with natural atoms.

A striking example of the power of circuit QED in the strong-coupling regime is the engineering of photonic states in the resonator demonstrated in 2009~\cite{Hofheinz2009}. In this experiment, a phase qubit (see \secref{sec:PhaseQubit}) was coupled to a transmission-line resonator. Through external control lines, it was possible to both tune the qubit frequency and to rotate the qubit state on the Bloch sphere. Starting with no photons in the resonator and $\omega_{01}$ detuned from $\omega_{\rm r}$, some qubit superposition state is prepared by a rotation. The qubit is then tuned into resonance with the resonator for a certain time, realizing a SWAP operation between the two systems. The qubit is then detuned again, prepared in another state through a rotation, tuned into resonance again to transfer part of its new state, and so on. Repeating this procedure, any photonic superposition state can be created in the resonator~\cite{Law1996, Liu2004}. The scheme relies on two particular strengths of circuit QED: tunable qubit frequency and strong coupling (to have enough time to perform all operations before coherence and/or energy is lost).

Other examples of circuit-QED experiments are too numerous to list here. Instead, we note that Josephson-junction qubits can be coupled not only to resonators with single modes, but also to open transmission-line \textit{waveguides} that support a continuum of modes. Experiments with such 1D waveguide-QED systems have yielded clear demonstrations of classical quantum-optics effects like the Mollow triplet~\cite{Astafiev2010} (the fluorescence from a driven qubit has peaks at three frequencies~\cite{Mollow1969}), perfect reflection of a single photon by a single qubit~\cite{Astafiev2010, Hoi2011}, and large cross-Kerr interaction between single photons (mediated by the two lowest transitions in a transmon qubit)~\cite{Hoi2013a}.


\subsection{New coupling strengths}

In cavity QED, the normalized coupling strength $\eta = g / \omega_{\rm r}$ usually does not become much larger than $10^{-6}$. The fundamental reason for this is that the fine structure constant $\alpha \approx 1 / 137$ is so small; a calculation shows that $g \propto \alpha^{3/2}$ in cavity-QED setups. However, surprisingly, $g$ scales differently with $\alpha$ in circuit QED~\cite{Devoret2007}. For the capacitive coupling between a transmon and an $LC$ resonator shown in \figref{fig:CircuitQED}, $g \propto \alpha^{1/2}$, and for a Josephson-junction qubit interrupting a transmission-line resonator, the scaling is $g \propto \alpha^{-1/2}$. Taking advantage of these favorable conditions for large coupling strengths, flux qubits have demonstrated first \textit{ultrastrong coupling} (USC; $\eta > 0.1$)~\cite{Niemczyk2010} and recently even \textit{deep strong coupling} ($\eta > 1$)~\cite{Yoshihara2017}.

When the light-matter coupling becomes ultrastrong, the Jaynes--Cummings Hamiltonian in \eqref{eq:JC} is no longer sufficient to describe the system. Instead, it is necessary to use the full quantum Rabi Hamiltonian
\be
H_{\rm Rabi} = \hbar \omega_{\rm r} a^\dag a + \hbar \omega_{01} \sz + \hbar g \mleft( a^\dag + a \mright) \mleft( \sp + \sm \mright).
\label{eq:Rabi}
\ee
The terms $a^\dag \sp$ and $a \sm$ can, for small $g$, be dropped using the so-called rotating-wave approximation, since they rotate rapidly (in the interaction picture) and average out on relevant timescales (set by $g$). For this reason, $a^\dag \sp$ and $a \sm$ are sometimes referred to as counter-rotating terms.

The inclusion of the counter-rotating terms breaks conservation of the number of excitations $N = a^\dag a + \sp \sm$ in the system, since $\comm{N}{H_{\rm JC}} = 0$ but $\comm{N}{H_{\rm Rabi}} \neq 0$. This makes it considerably more difficult to solve the quantum Rabi Hamiltonian analytically~\cite{Braak2011}. However, it also makes for more interesting physics in the system. One example is that the ground state of the Jaynes--Cummings Hamiltonian is the separable state with the qubit in $\ket{0}$ and no photons in the resonator, but the ground state of the quantum Rabi Hamiltonian contains excitations in both the qubit and the resonator. These ground-state excitations are \textit{virtual}; they are bound to the system and cannot escape, since a system in its ground state cannot lose energy.

Another interesting effect of the counter-rotating terms is that they allow \textit{higher-order processes} that do not conserve $N$. For example, if the qubit energy equals that of three photons in the resonator, a third-order process connects the system state with zero photons and the qubit in $\ket{1}$ with the system state that has three photons and the qubit in $\ket{0}$. On resonance, this third-order process creates a coherent coupling between these two states such that the system can oscillate directly between the two~\cite{Garziano2015}. Many more of these higher-order processes can be found, and the effective coupling strength $g_{\rm eff}$ for them can, although it is much smaller than $g$, be strong in a circuit-QED system. This means that various analogues of nonlinear optics can be realized~\cite{Kockum2017a, Stassi2017}.

This is but a small sample of the fascinating physics that takes place with ultrastrong light-matter coupling. For a more detailed review of USC in circuit QED, see Ref.~\cite{Gu2017}. For a recent general review of USC between light and matter, see Ref.~\cite{Kockum2019}.


\subsection{New selection rules}

When Josephson-junction qubits interact with electromagnetic fields, the interaction can cause transitions between different qubit states. In similar situations with natural atoms, the atomic eigenstates and the dipole moment (which gives the interaction with electromagnetic field) have well-defined parities, which gives rise to \textit{selection rules} for atomic transitions. The dipole moment has odd parity and can thus only cause transitions between atomic states of different parities, since symmetry considerations show that the matrix elements for transitions between states of the same parity would be zero~\cite{Cohen-Tannoudji1977}.

As we saw in \secref{sec:ThreeQubits}, all three basic Josephson-junction qubits have eigenstates that \textit{lack} well-defined parities. The only exceptions to this state of affairs are the sweet spots for charge and flux qubits, i.e., the points with half-integer background charge $n_{\rm g}$ for charge qubits and with half-integer normalized external magnetic flux $f$ for flux qubits. It is thus possible to \textit{control} the selection rules for interaction between Josephson-junction qubits and electromagnetic fields by simply varying an external control parameter.

\begin{figure}
\centering
\includegraphics[width=\linewidth]{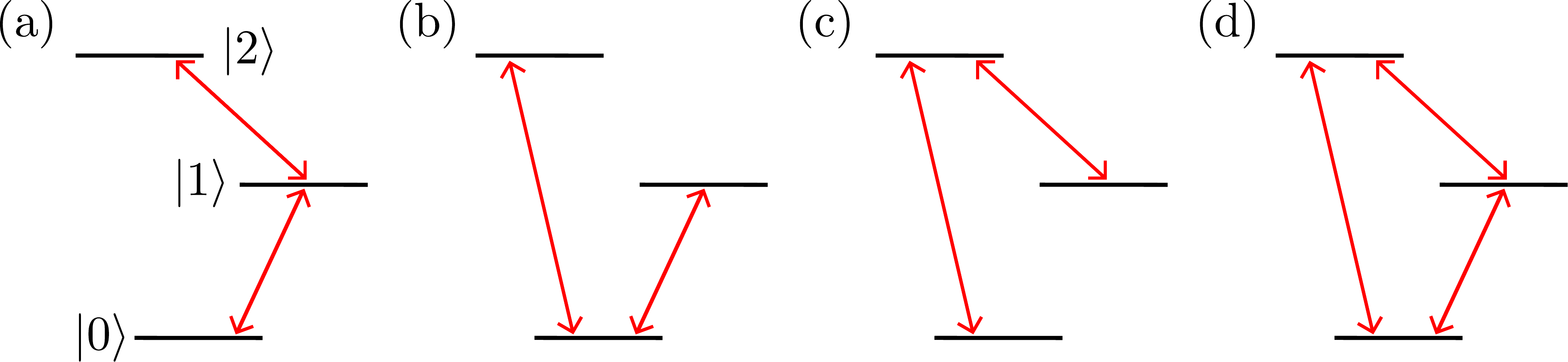}
\caption{Qutrits categorized by the allowed transitions between the system eigenstates.
(a) $\Xi$-type qutrit.
(b) $V$-type qutrit.
(c) $\Lambda$-type qutrit.
(d) $\Delta$-type qutrit. This last configuration is not possible for natural atoms.
\label{fig:ThreeLevelSystems}}
\end{figure}

The lack of selection rules for some Josephson-junction qubits becomes particularly interesting when we consider transitions also to the second excited state of the qubit, i.e., when we have a \textit{qutrit}. For natural three-level atoms, selection rules limit the possible level configurations and transitions to those shown in \figref{fig:ThreeLevelSystems}(a)-(c). However, a Josephson-junction qutrit, tuned such that it lacks selection rules, can also have transitions in the ``$\Delta$-type'' configuration shown in \figref{fig:ThreeLevelSystems}(d)~\cite{Liu2005}. The coexistence of all the three transitions shown has been confirmed in a flux-qubit experiment~\cite{Deppe2008}.

The existence of $\Delta$-type Josephson-junction qutrit enables several interesting applications. By driving the $\ket{0} \leftrightarrow \ket{2}$ transition in a system where the energy-relaxation rate for the $\ket{1} \leftrightarrow \ket{2}$ transition is fast, population inversion between $\ket{0}$ and $\ket{1}$ can be achieved. In this way, a weak probe at frequency $\omega_{01}$ can be amplified, which has been demonstrated in an experiment with a single flux-qubit in a waveguide~\cite{Astafiev2010a}. Similarly, the $\Delta$-type configuration makes frequency conversion possible~\cite{Liu2014, Sanchez-Burillo2016}. Frequency up-conversion occurs when photons are absorbed at $\omega_{01}$ and $\omega_{12}$, and a photon is emitted at $\omega_{02}$. Conversely, frequency down-conversion occurs when a photon is absorbed at $\omega_{02}$ and photons are emitted at $\omega_{01}$ and $\omega_{12}$. For more applications, see Ref.~\cite{Gu2017}.


\subsection{New atom sizes}

A standard assumption in quantum optics is that the atoms are \textit{small} compared to the wavelength of the light they interact with. This is certainly true for natural atoms (radius $r \approx \unit[10^{-10}]{m}$) coupled to optical light (wavelength $\lambda \approx \unit[10^{-6} - 10^{-7}]{m}$). Josephson-junction qubits are much larger: their size can reach \unit[$10^{-4} - 10^{-3}$]{m}. However, this is still small compared to the wavelength of the microwaves that couple to the qubits: $\lambda \approx \unit[10^{-2} - 10^{-1}]{m}$. Until recently, natural and artificial atoms alike were therefore routinely approximated as \textit{point-like} when calculating light-matter interaction; this is sometimes called the \textit{dipole approximation}.

In 2014, an experiment~\cite{Gustafsson2014} was performed that demonstrated coupling between a transmon qubit and propagating surface acoustic waves (SAWs), i.e., vibrations (\textit{phonons}) confined to the surface of a substrate~\cite{Morgan2007}. Since the SAWs propagated on a piezoelectric substrate (GaAs), the vibrations had an electromagnetic component that induced charge on the fingers of the large interdigitated capacitance shunting the SQUID in the transmon. The crucial point here is that the SAW phonons propagate at roughly the \textit{speed of sound}, while microwave photons in circuit QED propagate at almost the \textit{speed of light}. Since the SAWs are at microwave frequencies, this means that their wavelength is in the range $\lambda \approx \unit[10^{-7} - 10^{-6}]{m}$, which is clearly smaller than the size of the transmon qubit. Indeed, the distance between each finger in the interdigitated transmon capacitance was $\lambda/4$ in the experiment.

\begin{figure}
\centering
\includegraphics[width=0.7\linewidth]{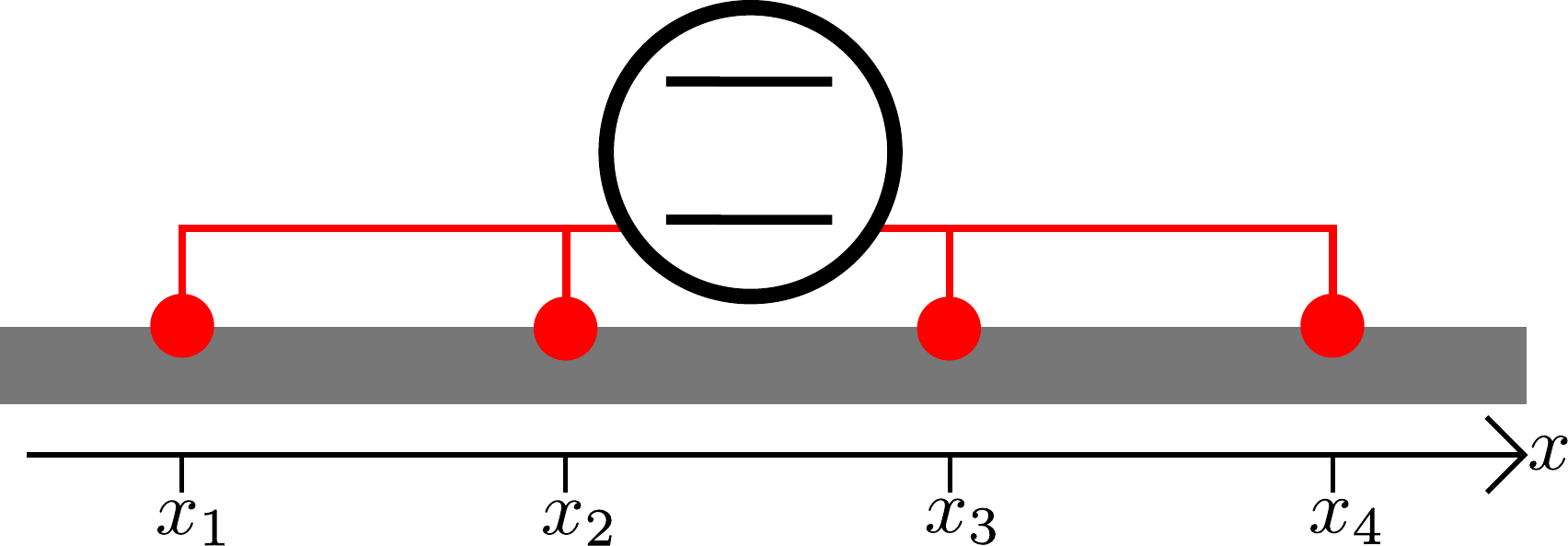}
\caption{Sketch of a giant atom coupled to an open waveguide (grey) at four points (red dots) where the distances between connection point coordinates $x_j$ is on the order of, or larger than, the wavelength of the waves propagating in the waveguide.
\label{fig:GiantAtom}}
\end{figure}

In setups with SAWs coupled to Josephson-junction qubits, it is thus justified to speak of ``\textit{giant atoms}'', atoms that couple to a bosonic field at multiple points, separated by wavelength distances, as sketched in \figref{fig:GiantAtom}. This introduces two main complications compared to the ``small-atom'' situation in standard quantum optics:
\begin{itemize}
\item The emission and absorption of excitations at the multiple connection points gives rise to new \textit{interference effects}~\cite{Kockum2014}. For example, a giant atom with two connection points spaced $\lambda/2$ apart is protected from decaying into the waveguide, since the emission from the two points will interfere destructively. Since $\lambda$ is set by the transition frequency of the atom, the energy-relaxation rate of the atom acquires a \textit{frequency dependence}, which is particularly interesting for Josephson-junction qubits that have tunable transition frequencies. This phenomenon can be used to protect quantum information from decoherence or to design situations where different transitions in a multi-level atom couple to the waveguide with different strengths.
\item The time it takes for excitations to travel from one connection point to the next can be non-negligible compared to the timescales of the atom dynamics~\cite{Guo2017}.
\end{itemize}

The field of circuit QAD (quantum acoustodynamics; the interaction between qubits and phonons)~\cite{Aref2016, Manenti2017} is now attracting much interest, but we note that giant atoms can be realized in a more conventional circuit-QED setup. One simply couples a Josephson-junction qubit to a transmission line, meander the line away on the chip until a wavelength distance has been reached, and then bring the waveguide back to couple to the qubit once more~\cite{Kockum2014}. In such a setup, interference effects for one and multiple~\cite{Kockum2018} giant atoms can be designed with greater precision than if SAWs are used.

\section{Acknowledgements}

We thank Xiu Gu, Adam Miranowicz, and Yu-xi Liu for useful discussions. We also thank Sergey Shevchenko and Roberto Stassi for helpful comments on drafts of this chapter. We acknowledge support from the MURI Center for Dynamic Magneto-Optics via the Air Force Office of Scientific Research (AFOSR) award No.~FA9550-14-1-0040, the Army Research Office (ARO) under grant No.~W911NF-18-1-0358, the Asian Office of Aerospace Research and Development (AOARD) grant No.~FA2386-18-1-4045, the Japan Science and Technology Agency (JST) [through the Q-LEAP program, the ImPACT program, and CREST Grant No.~JPMJCR1676], the Japan Society for the Promotion of Science (JSPS) through the JSPS-RFBR grant No.~17-52-50023 and the JSPS-FWO grant No.~VS.059.18N, the RIKEN-AIST Challenge Research Fund, and the John Templeton Foundation. 
 

\bibliography{JJQubitReferences}

\end{document}